\documentclass[%
reprint,
 amsmath,amssymb,
prb,
floatfix,
]{revtex4-2}

\usepackage{graphicx}
\usepackage{dcolumn}
\usepackage{bm}
\usepackage[dvipsnames]{xcolor}

\begin{document}

\preprint{PRB}

\title{Effective narrow ladder model for two quantum wires on a semiconducting substrate}

\author{Anas Abdelwahab}
 \affiliation{Leibniz Universit\"{a}t Hannover, Institut f\"{u}r Theoretische Physik, Appelstr.~2, 30167 Hannover, Germany}
\author{Eric Jeckelmann}
\affiliation{Leibniz Universit\"{a}t Hannover, Institut f\"{u}r Theoretische Physik, Appelstr.~2, 30167 Hannover, Germany}%

\date{\today}

\begin{abstract}
We present a theoretical study of two spinless fermion wires coupled to a three dimensional semiconducting substrate. We develop a mapping of wires and substrate onto a system of two coupled two-dimensional ladder lattices using a block Lanczos algorithm.
We then approximate the resulting system by narrow ladder models, which can be investigated using the density-matrix renormalization group method.
In the absence of any direct wire-wire hopping we find that the substrate can mediate an effective wire-wire coupling so that the wires could form an effective two-leg ladder with a Mott charge-density-wave insulating ground state for arbitrarily small nearest-neighbor repulsion. In other cases the wires remain effectively uncoupled even for strong wire-substrate hybridizations leading to the possible  stabilization of the Luttinger liquid phase at finite nearest-neighbor repulsion as found previously for single wires on substrates. These investigations show that it may be difficult to determine under which conditions the physics of correlated one-dimensional electrons can be realized in arrays of atomic wires on semiconducting substrates because they seem to 
depend on the model (and consequently material) particulars.
\end{abstract}

\maketitle

\section{\label{sec:intro}Introduction}
Systems of metallic atomic wires deposited on semiconducting substrates attracted lots of attention in the last two decades. One of the main issues regarding these systems is the existence of features related to one-dimensional (1D) electrons, e.g. Luttinger liquid phases~\cite{blum11,blum12,ohts15,yaji13,yaji16}, Peierls metal-insulator transitions and charge-density-wave states~\cite{yeom99,cheo15,shin12,aulb13}.
For instance, there is an ongoing debate about the existence of Luttinger liquid behavior in gold chains on the Ge(100) substrate~\cite{blum11,blum12,nak11,par14,jon16,sei16,sei18}. It is even disputed whether this material is effectively a quasi-one-dimensional system of weakly coupled chains~\cite{blum11,blum12} or an anisotropic two-dimensional system~\cite{nak11,par14,jon16}. Another system that reveals Luttinger liquid behavior is Bi deposited on InSb(001) surfaces in angle-resolved photoelectron spectroscopy~\cite{ohts15}. However, this behavior is observed for large coverage of Bi on the InSb substrate and thus it is also unclear whether the system
can be seen as made of separate atomic wires.
Thus a key question for all theses materials is whether there is a significant coupling between atomic wires.

However, the theoretical framework of 1D correlated electrons is derived primarily from purely 1D models~\cite{Schoenhammer,giamarchi07,solyom,gruener,chen}, which are then extended to anisotropic two (2D) and three dimensional (3D) systems. These extensions are not applicable for metallic atomic wires on semiconducting substrates due to their strong asymmetric nature, i.e. they represent arrays of 1D wires coupled to a 3D reservoir. Therefore, even if we assume that the atomic wires themselves are systems of 1D correlated electrons, it is necessary to investigate two aspects: firstly, the influence of the coupling to the 3D bulk semiconducting substrate on the 1D features; secondly, the possibility of substrate-mediated coupling between the wires.

\begin{figure}
\includegraphics[width=0.3\textwidth]{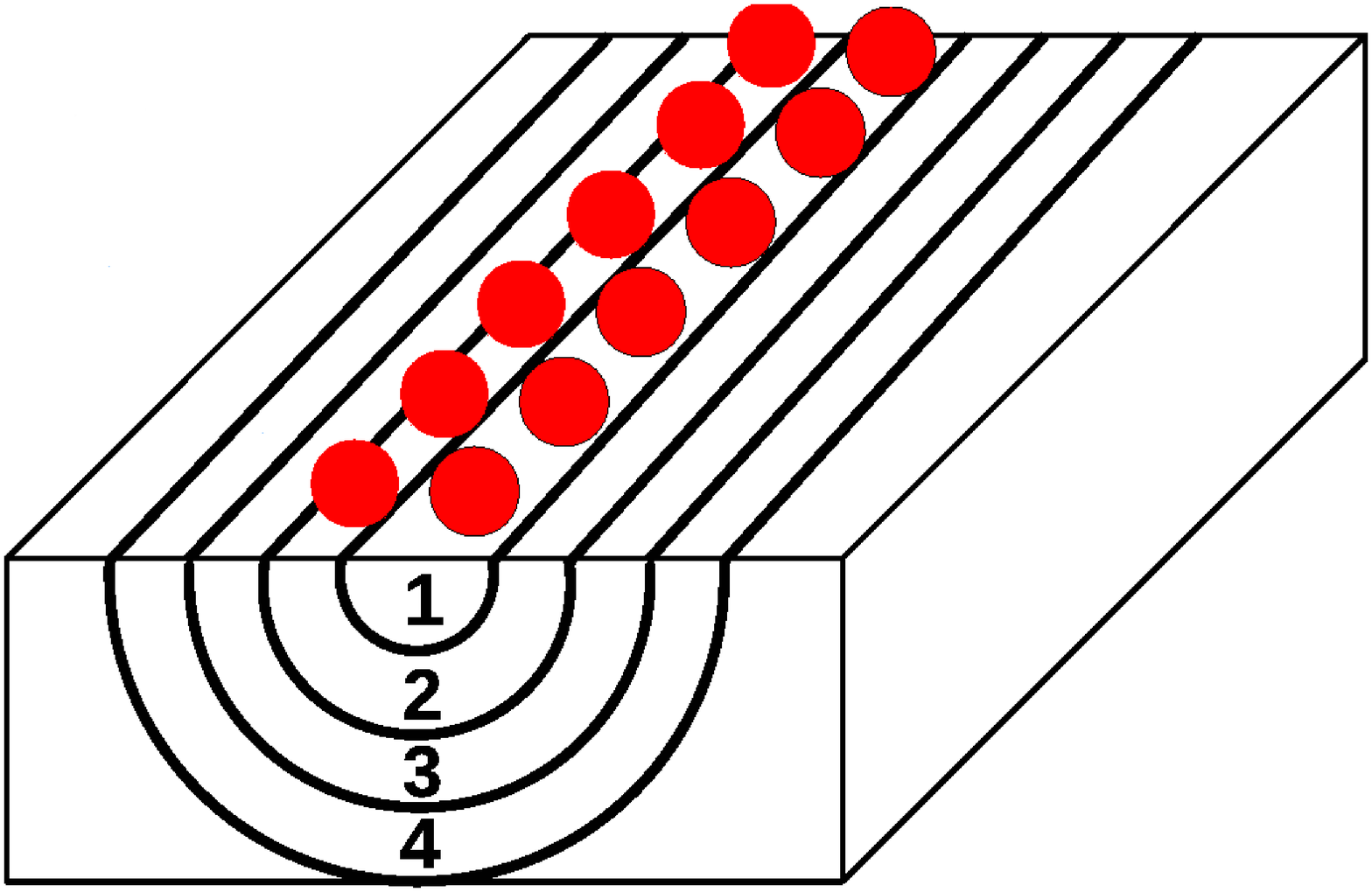}
\hspace*{2cm}
\includegraphics[width=0.3\textwidth]{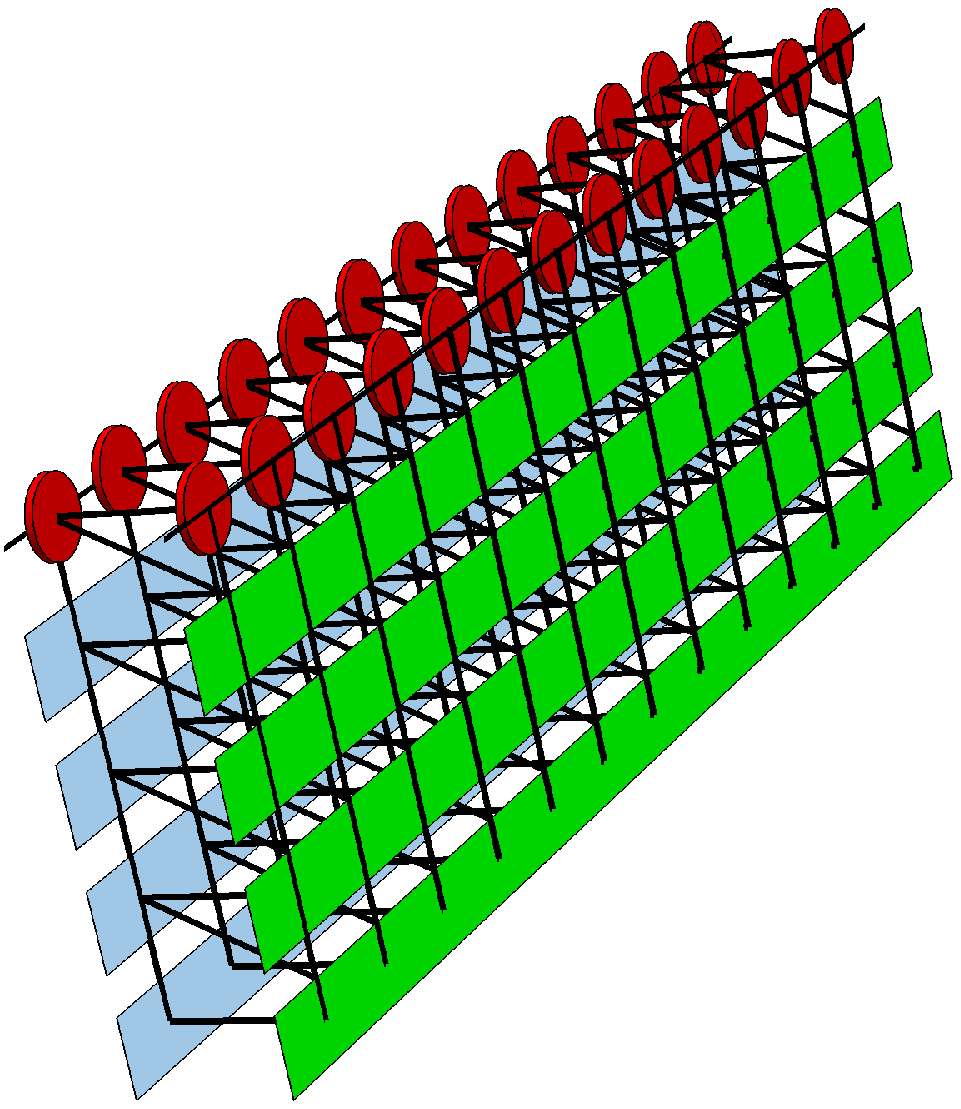}
\caption{\label{fig:mapping} 
The upper panel display sketch of two atomic wires (red spheres) on a 3D substrate with four numbered shells.
The lower panel ladder representation of the same system with the upper-most legs corresponding
to the atomic wires (red circles) and the other legs (in green and light blue) representing the shells one to four.
}
\end{figure}

In a series of previous publications~\cite{abd17a,abd17b,abd18}, we addressed the first issue. We established that, indeed, the coupling of a single metallic atomic wire to a 3D semiconducting substrate can stabilize the 1D nature of the wire and, in particular, can support the occurrence of Luttinger liquid phases. In the current article we would like to address the second issue. We develop a method to map multi wires coupled to semiconducting substrates onto a system of coupled 2D ladders (one ladder per wire). This method is based on the block Lanczos algorithm~\cite{shi14}. Then we approximate the original systems by keeping only a few legs of each 2D ladder, i.e. by constructing a narrow ladder model (NLM) that can be investigated using
well-established methods for quasi-one-dimensional correlated quantum systems. The original system and the resulting NLM are depicted in Fig.~\ref{fig:mapping}.

Using exact diagonalizations of noninteracting wires and the
density-matrix renormalization group (DMRG) method~\cite{whi92,whi93},
we investigate in details two wires on a semiconducting substrate (TWSS) for spinless fermions without direct wire-wire coupling and compare to the known results for a single wire on a substrate~\cite{abd18} and for two-leg ladders without substrate~\cite{don01}.
We find that the substrate can mediate an effective wire-wire coupling so that the atomic wires form an effective two-leg ladder that is known to have a Mott charge-density-wave insulating ground state for arbitrarily small nearest-neighbor repulsion. In other cases the wires remain effectively uncoupled even for strong wire-substrate hybridizations, which should result in a Luttinger liquid phase at finite nearest-neighbor repulsion as found previously for single wires on substrates.

The article is organized as follows. In the second section we introduce the TWSS model, the mapping of multi-wire-substrate models onto systems of coupled multi 2D ladders and the approximation by few-leg NLM. In the third section we discuss noninteracting wires while in the fourth section we present our results for correlated wires. We conclude in the fifth section.

\section{Modeling two wires on semiconducting substrates and the approximation by narrow ladder models}

In this section we construct a model for TWSS and approximate it by NLM. This procedure can be easily generalized to more than two wires on a semiconducting substrate.
Throughout our paper
we use different letters ($c,d,f$ and $g$) for the fermion operators
corresponding to different one-electron bases (representations),
and indices to distinguish the various one-electron states in a given basis.

\subsection{\label{sec:TWSS_model} The model of two wires on semiconducting substrate}

The substrate is described as explained in Ref.~\cite{abd17a} but since we focus on spinless fermion wires, we omit spin degrees of freedoms.
We restrict ourselves to insulating or semiconducting substrates, hence each substrate site has two orbitals, one contributing to the formation of the conduction bands and one to the valence bands.
Thus, the substrate Hamiltonian in real space takes the form
\begin{eqnarray}\label{eq:real-sub}
H_{\text{s}} &=& H_{\text{c}} + H_{\text{v}} \nonumber \\
&=& \sum_{\text{s}=\text{v},\text{c}}  \left( \epsilon_{\text{s}} \sum_{\mathbf{r}} n_{\text{s} \mathbf{r} } - t_{\text{s}} \sum_{\langle \mathbf{r} \mathbf{q} \rangle}  \left (
c^{\dag}_{\text{s} \mathbf{r}}  c^{\phantom{\dag}}_{\text{s}\mathbf{q}} + \text{H.c.}
\right ) \right).
\end{eqnarray}
The first sum runs over conduction ($\text{s=c}$) and valence ($\text{s=v}$) bands.
The second sum runs over all sites of a cubic lattice and the third one over all pairs $\langle \mathbf{r} \mathbf{q} \rangle$ of nearest-neighbor lattice sites.
The operator $c^{\dag}_{\text{s} \mathbf{r}}$ creates a spinless fermion on the orbital $\text{s}$ localized on the site with coordinates $\mathbf{r} = (x,y,z)$,
and the fermion density operator on each orbital is $n_{{\text s} \mathbf{r}} = c^{\dag}_{{\text s} \mathbf{r}} c^{\phantom{\dag}}_{{\text s} \mathbf{r} }$.
The transformation to the momentum-space is done only in $x$-direction which is the alignment direction of the two wires.
The other two dimensions are irrelevant for the two wires and they can remain in the real-space representation for the substrate.
This allows a mixed real-space momentum-space representation $\mathbf{r}_{k_x} = (k_x,y,z)$ where $k_x$ is the wave vector component in $x$-direction.
This representation is more convenient for the purpose of ladder mapping.
The Hamiltonian~(\ref{eq:real-sub}) takes the form
\begin{eqnarray}
\label{eq:mixed-sub}
H_{\text{s}} &=& \sum_{\text{s}=\text{v},\text{c}} \left [ \sum_{k_x,y,z} \epsilon_{\text{s}}(k_x) d^{\dag}_{{\text s} \mathbf{r}_{k_x}} d^{\phantom{\dag}}_{{\text s} \mathbf{r}_{k_x}} \right . \nonumber \\
&-& \left . t_{s} \sum_{k_x, \left \langle \left (y^{\phantom{\prime}},z^{\phantom{\prime}} \right), \left (y^{\prime},z^{\prime} \right) \right \rangle}
d^{\dag}_{{\text s} \mathbf{r}_{k_x}^{\phantom{\prime}}} d^{\phantom{\dag}}_{{\text s} \mathbf{r}_{k_x}^{\prime}} \right ]
\end{eqnarray}
with two single-electron dispersions 
\begin{equation}
\label{eq:mixed-valance-disp}
\epsilon_{\text{s}=\text{v},\text{c}}(k_x)  = \epsilon_{\text{s}=\text{v},\text{c}} - 2t_{\text{s}=\text{v},\text{c}} \cos(k_x),
\end{equation}
 the sum
over nearest-neighbor site pairs $\left \langle \left (y^{\phantom{\prime}},z^{\phantom{\prime}} \right), \left (y^{\prime},z^{\prime} \right) \right \rangle$
in the $yz$-layer for a given $k_x$, and $\mathbf{r}_{k_x}^{\prime} = (k_x,y^{\prime},z^{\prime})$.
The (possibly indirect) gap between the bottom of the conduction band and the top of the valence band is given by
$\Delta_{\text{s}} =  \epsilon_{\text{c}} - \epsilon_{\text{v}} - 6 \left (  \vert t_{\text{v}} \vert + \vert t_{\text{c}} \vert \right )$
and the condition $\Delta_{\text{s}} \geq 0$ restricts the range of allowed model parameters.

A simple representation of the wires is achieved using two 1D chains, possibly coupled by a single particle-hopping $t_{ab}$ between adjacent sites in the two wires.
The two wires are described by the Hamiltonian
\begin{eqnarray}
\label{eq:real-spinless-fermion_wires}
 H_{\text{w}} &=&
  \sum_{\text{w}=\text{a},\text{b}} \left( \epsilon_{\text{w}} \sum_{x} n_{x y_{\text{w}}} 
-t_{\text{w}}  \sum_{x} \left ( c^{\dag}_{x y_{\text{w}}}  
c^{\phantom{\dag}}_{x+1,y_{\text{w}}} + \text{H.c.} \right ) \right. \nonumber \\
&+& \left. V \sum_x n_{x y_{\text{w}}} n_{x+1 y_{\text{w}}} \right) 
-t_{\text{ab}}  \sum_{x} \left ( c^{\dag}_{x y_{\text{a}}}  
c^{\phantom{\dag}}_{x y_{\text{b}}} + \text{H.c.} \right ). \nonumber \\
\end{eqnarray}
The wires $\text{w} \equiv a$ and $b$ of length $L_x$ are aligned in the $x$-direction at positions $\mathbf{r} = (x,y_{\text{a}},0)$
and $\mathbf{r} = (x,y_{\text{b}},0)$ where $y_{\text{a}} \text{ and } y_{\text{b}} \in \{1,\dots,L_y\}$.
The sums over $x$ run over all wire sites.
The operator $c^{\dag}_{x y_{\text{a}}}$
creates an electron on the wire site $\mathbf{r} = (x,y_{\text{a}},0)$ for $\text{w} \equiv a$ and the operator $c^{\dag}_{x y_{\text{b}}}$ creates an electron on the wire site $\mathbf{r} = (x,y_{\text{b}},0)$ for $\text{w} \equiv b$.
Every site in the wires is exactly on top of the corresponding substrate site.
$t_{\text{a}}$ and $t_{\text{b}}$ are the usual hopping terms between
nearest-neighbor sites in the corresponding wire.
$\epsilon_{\text{a}}$ and $\epsilon_{\text{b}}$ are on-site potentials of the corresponding wires and $V$ is the interaction between fermions on nearest-neighbor sites.
The single-particle hopping $t_{\text{ab}}$ determines the direct coupling between the two wires.
The Hamiltonian of the two spinless-fermion wires can be written in the mixed representation 
\begin{eqnarray}
\label{eq:mixedrep-spinless-fermion-wires}
H_{\text{w}} & = &
 \sum_{\text{w}=\text{a},\text{b}} \sum_{k_x} \epsilon_{\text{w}}(k_x) d^{\dag}_{k_x y_{\text{w}}} d^{\phantom{\dag}}_{k_x y_{\text{w}}}   \nonumber \\
 &&+ \frac{V}{L_x}  \sum_{\text{w}=\text{a},\text{b}} \ \sum_{k_x,k^{\prime}_x,k^{\prime\prime}_x,k^{\prime\prime\prime}_x} 
 \left [ d^{\dag}_{k_x y_{\text{w}}} d^{\phantom{\dag}}_{k^{\prime}_x y_{\text{w}}}  d^{\dag}_{k^{\prime\prime}_x y_{\text{w}}} d^{\phantom{\dag}}_{k^{\prime\prime\prime}_x y_{\text{w}}} \right . \nonumber \\
 &&
 \left . \times \delta(k_x+k^{\prime\prime}_x-k^{\prime}_x-
 k^{\prime\prime\prime}_x) \right ] \nonumber \\
 && -t_{\text{ab}}  \sum_{k_x} \left ( d^{\dag}_{k_x y_{\text{a}}}  
d^{\phantom{\dag}}_{k_x y_{\text{b}}} + \text{H.c.} \right )
\end{eqnarray}
with two wire single-electron dispersions
\begin{equation}
\label{eq:wire-disp_a}
\epsilon_{\text{w}=\text{a},\text{b}}(k)  = \epsilon_{\text{w}=\text{a},\text{b}} - 2t_{\text{w}=\text{a},\text{b}} \cos(k) .
\end{equation}
Here $k_x,k^{\prime}_x,k^{\prime\prime}_x$, and $k^{\prime\prime\prime}_x$ denote momenta in the $x$-direction. 
$\delta(k) = 1$ if $k \mod 2 \pi = 0$
and $\delta(k)=0$ otherwise.

The hybridization between each wire and the substrate is modeled by 
\begin{equation}
\label{eq:hybridization-insulator}
H_{\text{ws}} = H_{\text{wa}} + H_{\text{wb}} =
 \sum_{\text{w}=\text{a},\text{b}; \text{s}=\text{v},\text{c}} (-t_{\text{ws}}) \sum_{x} \left ( c^{\dag}_{{\text{s}} \mathbf{r}_{\text{w}}}  
c^{\phantom{\dag}}_{x y_{\text{w}}} + \text{H.c.}  \right )
\end{equation}
which represents a hopping between each wire site and the nearest valence and conduction band sites at $\mathbf{r}_{\text{w}} = (x,y_{\text{w}},1)$, w=a,b.
In the mixed representation the wire-substrate hybridization becomes
\begin{eqnarray}
\label{eq:hybridization-insulator-mixed-rep}
H_{\text{ws}} & = & H_{\text{wa}} + H_{\text{wb}} \nonumber \\
& = &\sum_{\text{w}=\text{a},\text{b}; \text{s}=\text{v},\text{c}} (-t_{\text{ws}}) \sum_{k_x} \left ( d^{\dag}_{{\text{s}} \mathbf{k}_{\text{w}}}  
d^{\phantom{\dag}}_{k_x y_{\text{w}}} + \text{H.c.}  \right )
\end{eqnarray}
with $\mathbf{k}_{\text{w}} = (k_x,y_{\text{w}},1)$, w=a,b.
Thus the full TWSS Hamiltonian is given by
\begin{equation}\label{eq:TWS_model}
 H =  H_{\text{w}} + H_{\text{s}} + H_{\text{ws}.} 
\end{equation}

\subsection{Two-impurity  subsystems\label{sec:two-impurity}}
In this section we reformulate the Hamiltonian~(\ref{eq:TWS_model}) as a set of two-impurity-host subsystems.
This is first done for the noninteracting case ($V=0$). In the mixed representation, the Hamiltonian takes the form 
\begin{equation}
\label{eq:hamiltonianWSkx}
 H=\sum_{k_x} H_{k_x}
\end{equation}
where $H_{k_x}$ are independent sub-system Hamiltonians such that $\left[H_{k_x},H_{k_x^{\prime}}\right]=0$ $\forall k_x,k_x^{\prime}$.
Each sub-system Hamiltonian represents a two-impurity subsystem that takes the form
\begin{eqnarray}
H_{k_x}&=& \epsilon_{\text{a}}(k_x) \, d^{\dag}_{k_x y_{\text{a}}}
d^{\phantom{\dag}}_{k_x y_{\text{a}}} 
+ \epsilon_{\text{b}}(k_x) \, d^{\dag}_{k_x y_{\text{b}}}
d^{\phantom{\dag}}_{k_x y_{\text{b}}} \nonumber \\
&-& t_{\text{ab}}  \left ( d^{\dag}_{k_x y_{\text{a}}}  
d^{\phantom{\dag}}_{k_x y_{\text{b}}} + \text{H.c.} \right ) \nonumber \\ 
&+&  
\sum_{\text{s}=\text{v},\text{c}} \left [ \sum_{y,z} \epsilon_{\text{s}}(k_x) d^{\dag}_{{\text s} \mathbf{r}_{k_x}} d^{\phantom{\dag}}_{{\text s} \mathbf{r}_{k_x}} \right . \nonumber \\
&-& \left . t_{s} \sum_{\left \langle \left (y^{\phantom{\prime}},z^{\phantom{\prime}} \right), \left (y^{\prime},z^{\prime} \right) \right \rangle}
d^{\dag}_{{\text s} \mathbf{r}_{k_x}^{\phantom{\prime}}} d^{\phantom{\dag}}_{{\text s} \mathbf{r}_{k_x}^{\prime}} \right ]
 \nonumber \\
&-& \sum_{\text{s}=\text{v},\text{c};\text{w}=\text{a},\text{b}} \left ( t_{\text{ws}} d^{\dag}_{{\text{s}} \mathbf{k}_{\text{w}}}  
d^{\phantom{\dag}}_{k_x y_{\text{w}}} + \text{H.c.} \right ) .
\label{eq:impurity-insulator}
\end{eqnarray}
$H_{k_x}$ represents two non-magnetic impurities with the energy levels
$\epsilon_{\text{a}} (k_x)$ and $\epsilon_{\text{b}} (k_x)$ corresponding to the two wires.
These energy levels are coupled to a 2D host determined by a substrate ($y,z$)-slice through
the hybridization parameter $t_{\text{ws}}$.
Each ($y,z$)-slice corresponds only to the given wave vector $k_x$.
For a noninteracting wire, each $H_{k_x}$
is a single-particle problem and it is amenable to exact diagonalization.
Each single-particle Hamiltonian has the dimension $N_{\text{imp}}=2L_yL_z+2$. 

\subsection{Ladder representation \label{sec:ladder}}
The two-impurity subsystem can be mapped onto a two-leg ladder system using the Block-Lanczos (BL) algorithm. This method has been used to investigate 
multiple quantum impurities embedded in 
multi-dimensional noninteracting hosts~\cite{shi14,Allerdt2015,Allerdt2017}
and, very recently, a single wire coupled to two multi-dimensional noninteracting leads~\cite{Lange2020}.
The BL algorithm is an extension of the Lanczos algorithm to formulate block tridiagonal matrix starting from more than one basis state. The number of basis states chosen to start the iterations determines the size of each single block within the resulting block tridiagonal matrix.

The BL procedure starts with a matrix $P_1$ with  $N_{\text{row}}=N_{\text{imp}}$ rows and $N_{\text{col}}$ columns. The row index numbers the single-electron basis states  $({\text{s}} \mathbf{r}_{k_x})$
in the mixed representation for a given $k_x$
while the column index numbers
the impurities (the wires). Here we have $N_{\text{col}}=2$ but if we had more than two wires, the multi-impurity subsystem would have more than two impurities and $N_{\text{col}}>2$ would be correspondingly larger.
The first column of $P_1$ is $(1, 0, 0, \dots, 0)$ and
corresponds to the single-electron state
 $d^{\dag}_{k_x y_{\text{a}}}| \Phi \rangle$ 
 representing the first impurity site 
 while the second column of $P_1$ is $(0, 1, 0, \dots, 0)$
 and corresponds to the single-electron state
 $d^{\dag}_{k_x y_{\text{b}}}| \Phi \rangle$ 
 representing the second impurity site
 ($| \Phi \rangle$ is the vacuum state).
The BL iteration is defined as
\begin{equation}
 \label{eq:BL_iteration}
 P_{l+1} T^{\dag}_l = H_{k_x} P_l - P_l E_l - P_{l-1} T_{l-1}
\end{equation}
where $E_l=P^{\dag}_l H_{k_x} P_l$, $P_0=0$ and $T_0=0$.
The decomposition of the left-hand side of~(\ref{eq:BL_iteration}) into two matrices
$P_{l+1}$ and  $T^{\dag}_l$ can be obtained using the $\textbf{QR}$ decomposition~\cite{recipes}.
Thus $P_l$ is a column-orthogonal matrix and $T_l$ is a lower-triangular matrix, i.e. with matrix elements $\left[\tau_{l}\right]_{n,n'}=0 $ for $n<n'$.
We denote the matrix elements of $E_l$ as $\left[e_l\right]_{n,n'}$.
The BL basis $P_l$ spans the Krylov subspace of $H_{k_x}$.
The full implementation of the BL method generates a matrix 
\begin{equation}
 \label{eq:matix_P}
  \textbf{P}=\left[P_1 \hspace{2mm} P_2 \hspace{2mm} P_3 \hspace{2mm} ...\right],
\end{equation}
which has the dimensions $N_{\text{row}}=N_{\text{col}}=N_{\text{imp}}$.
The matrix $\textbf{P}$ can be used to block-tridiagonalize the Hamiltonian $H^{BL}_{k_x}$ in the form
\begin{equation}
\label{eq:BL2legMatrix}
 H^{BL}_{k_x} = \begin{bmatrix}
               E_1 & T_1 & 0 & 0 & \cdots \\
               T^{\dag}_{1} & E_2 & T_2 & 0 & \cdots \\
               0 & T^{\dag}_{2} & E_3 & T_3 & \cdots \\
               0 & 0 & T^{\dagger}_{3} & E_4 & \ddots \\
               \vdots & \vdots & \vdots & \ddots & \ddots
              \end{bmatrix} .
\end{equation}
Since the number of impurities is two, each block in~(\ref{eq:BL2legMatrix}) is a $2\times2$ matrix.
Thus, $H^{BL}_{k_x}$ represents a two-leg ladder which is written in the form
\begin{eqnarray}
\label{eq:hamiltonianBL2Imp}
H^{BL}_{k_x}&=& 
\sum^{N_{\text{imp}}/2}_{l=1}\sum^{2}_{n,n'=1}
\left[e_l(k_x)\right]_{n,n'}f^{\dag}_{k_x ln}f^{\phantom{\dag}}_{k_x ln'} \\
&+&\sum^{\left[N_{\text{imp}}/2\right]-1}_{l=1}\sum^{2}_{n,n'=1} \left [
\left[\tau_{l}(k_x)\right]_{n,n'} f^{\dag}_{k_xln\phantom{(}} f^{\phantom{\dag}}_{k_x,l+1,n'} 
+\text{H.c.} \right] \nonumber
\end{eqnarray}
where 
the new fermion operators $f_{k_x,l,n}$
are given by the columns of the matrices $P_l$. More precisely, 
\begin{eqnarray}
f_{k_x,1,1} &=& d_{k_x y_{\text{a}}} \\\ 
f_{k_x,1,2} &=& d_{k_x y_{\text{b}}} 
\end{eqnarray}
and
\begin{equation}
f_{k_x,l>1,n} = \sum_{m} [P_l]_{m,n}
d_{{\text{s}} \mathbf{r}_{k_x}}
\end{equation}
where the row index $m$ numbers the $N_{\text{imp}}$
basis states $({\text{s}} \mathbf{r}_{k_x})$ 
in the mixed representation for a given $k_x$.
In practice, we never calculate the operators $f^{\phantom{\dag}}_{k_x,l,n}$ as only the Hamiltonian
matrix elements $\left[e_l(k_x)\right]_{n,n'}$ and $\left [\tau_{l}(k_x)\right]_{n,n'}$ are required for our method.

We distinguish two kinds of wire-substrate hybridizations.
The first kind is the hybridization of the two wires with two substrate sites belonging to different sublattices in the substrate bipartite lattice, e.g. the wires are nearest neighbor (NN) with $|y_{\text{a}}-y_{\text{b}}|=1$.
In this case the system is particle-hole symmetric, i.e. the Hamiltonian is invariant under the transformation $f^{\phantom{\dag}}_{k_x ln} \rightarrow (-1)^l (-1)^n f^{\dag}_{k_x ln}$, and thus half-filling corresponds to the Fermi energy $\epsilon_{\text{F}}=0$.
The second kind is the hybridization of the two wires with two substrate sites belonging to the same sublattice in the substrate bipartite lattice, e.g. the wires are next nearest neighbor (NNN) with $|y_{\text{a}}-y_{\text{b}}|=2$.
In this case the system is not particle-hole symmetric as long as $t_{\text{ab}}\neq 0$ and the full wire-substrate lattice is not bipartite.
However, for $t_{\text{ab}}=0$ the system is again bipartite and particle-hole symmetric.

\begin{figure}[t]
\includegraphics[width=0.30\textwidth]{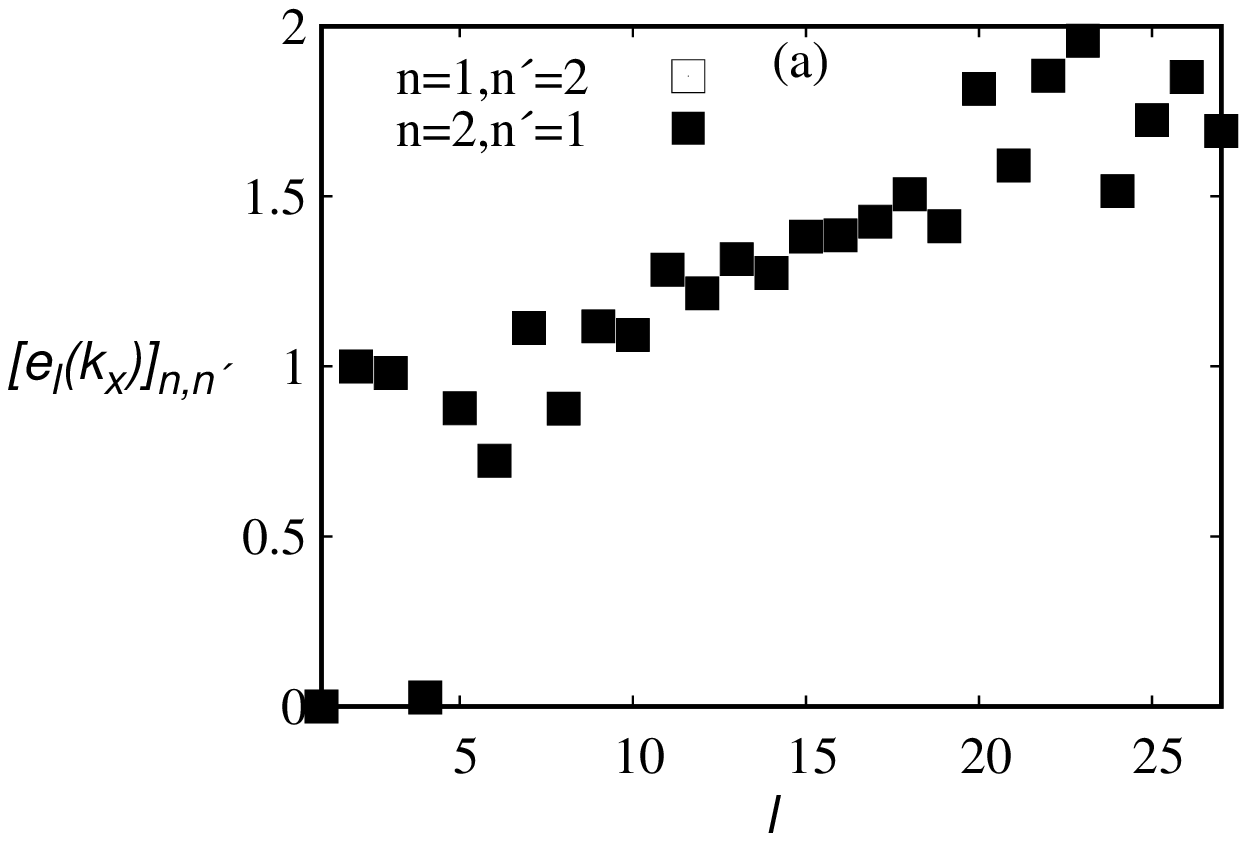}
\includegraphics[width=0.30\textwidth]{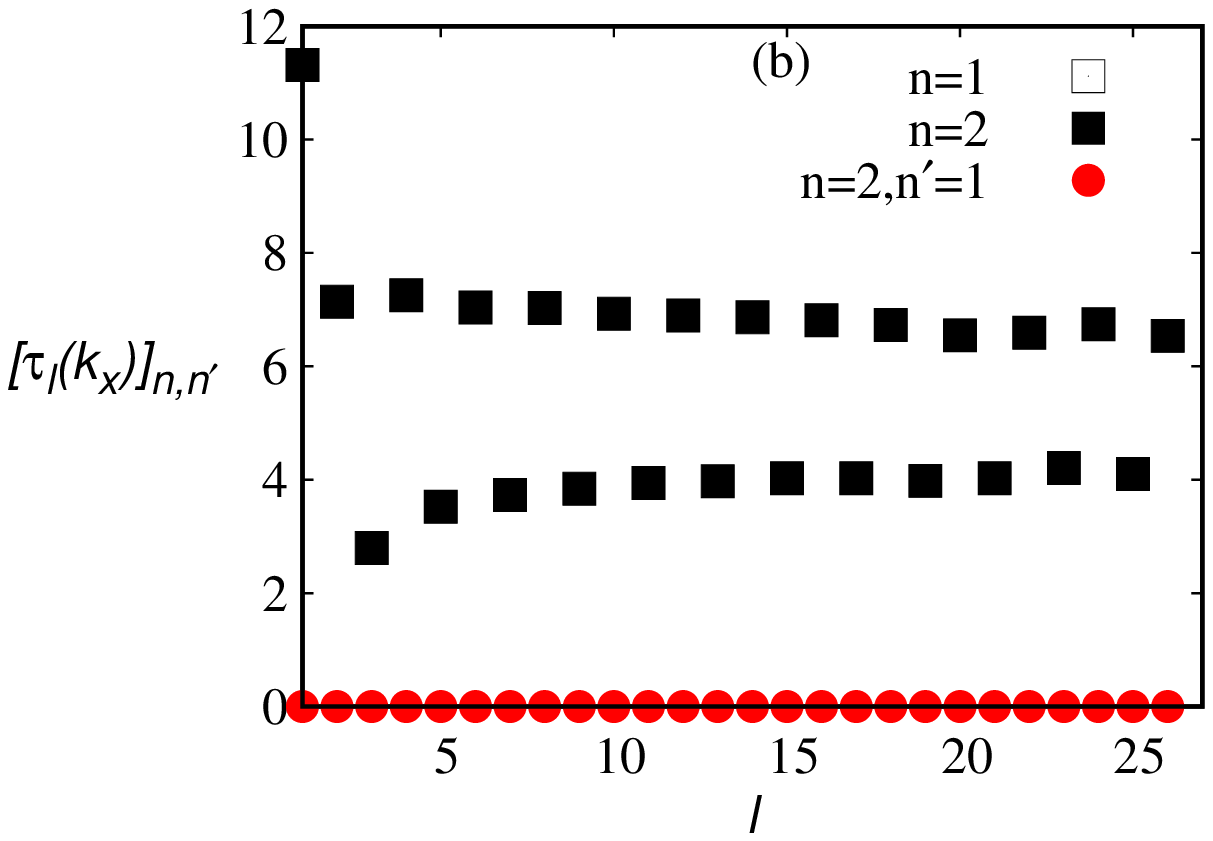}
\includegraphics[width=0.30\textwidth]{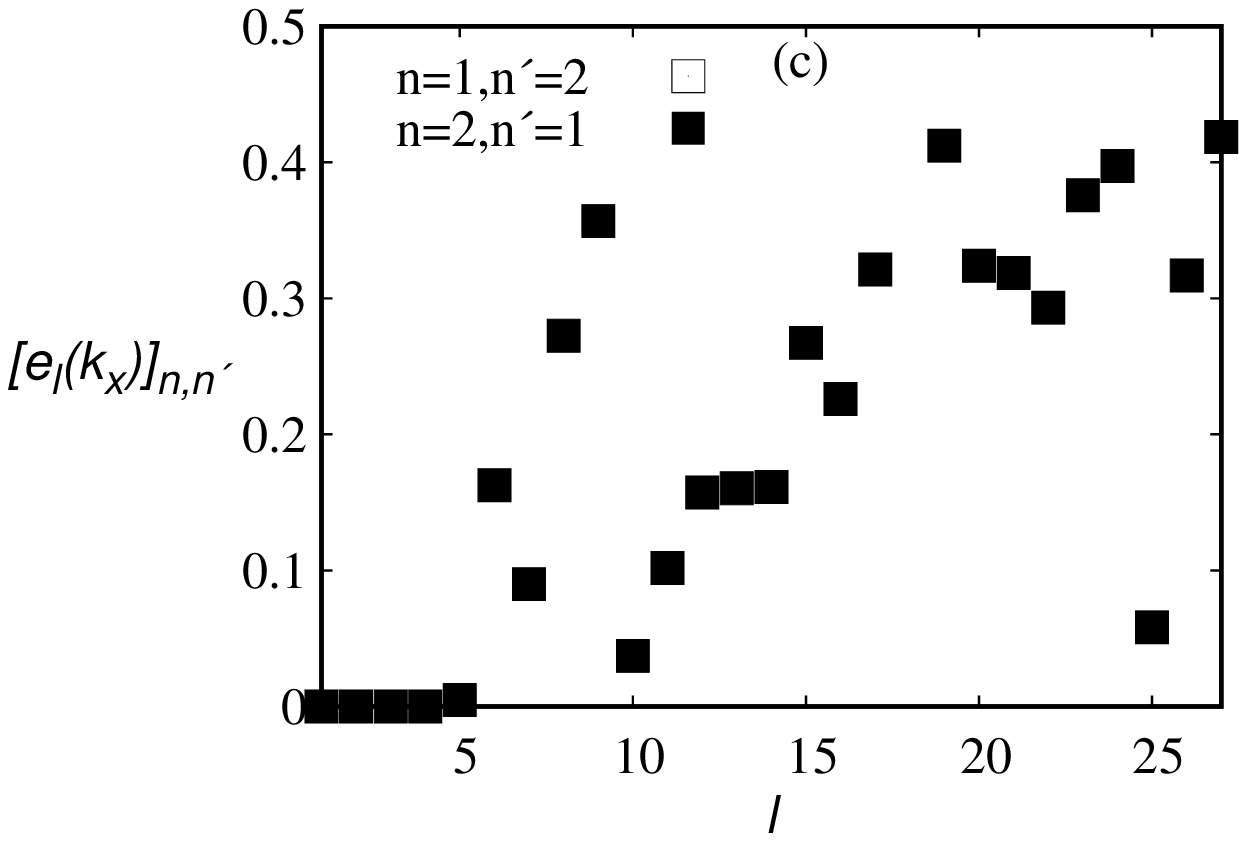}
\includegraphics[width=0.30\textwidth]{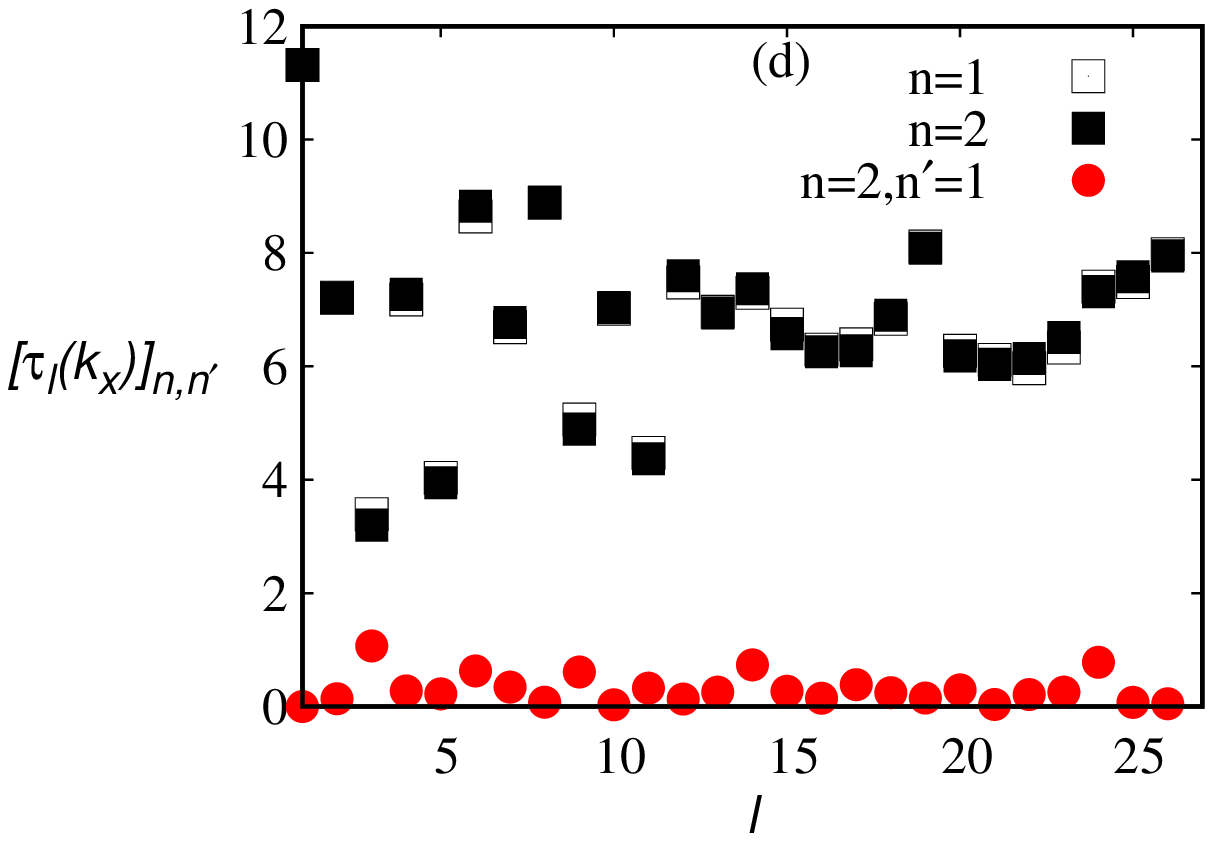}
\caption{\label{fig:paramtw3tww0ts1tws8} 
Hopping parameters of the two-impurity subsystem after the Block-Lanczos transformation. The parameters are given in the text. (a) Intra-rung hopping terms $\left[e_l(k_x)\right]_{n,n'}$ for the system with NN impurities.
(b) Intra-leg hopping terms $\left[\tau_{l}(k_x)\right]_{n,n}$ as well as inter-leg diagonal hoppings $\left[\tau_{l}(k_x)\right]_{1,2}$ for the system with NN impurities.
(c) Intra-rung hopping terms $\left[e_l(k_x)\right]_{n,n'}$ for the system with NNN impurities.
(d) Intra-leg hopping terms $\left[\tau_{l}(k_x)\right]_{n,n}$ as well as inter-leg diagonal hoppings $\left[\tau_{l}(k_x)\right]_{1,2}$ for the system with NNN impurities. The horizontal axis represents the number of Bloc-Lanczos shell. 
Note that the results are indistinguishable for n=1 and n=2 (square symbols) in (a), (b) and (c) but not  in (d).
}
\end{figure}

To illustrate the procedure, we calculate the parameters $\left[e_l(k_x)\right]_{n,n'}$ and $\left[\tau_{l}(k_x)\right]_{n,n'}$  for NN impurities coupled to an insulating substrate with the wire-substrate model parameters $t_{\text{w}}=3$, 
$t_{\text{ab}}=0$, $t_{\text{s}}=1$ and $t_{\text{ws}}=8$,
i.e. without direct coupling between the two impurities.
The diagonal terms $\left[e_l(k_x)\right]_{n,n}$ depend of the dispersion in the wire direction and the on-site chemical potentials.
For $n\neq n'$ the inter-leg hopping terms $\left[e_l(k_x)\right]_{n,n'}$  vary as $l$ increases as shown in Fig.~\ref{fig:paramtw3tww0ts1tws8}(a).
Since the initial vectors are associated with sites belonging to different sublattices,  the inter-rung hopping terms are $\left[\tau_{l}(k_x)\right]_{n,n'}=0$ for $n\neq n'$ (the BL algorithm always enforces $\left[\tau_{l}\right]_{n,n'}=0 $ for $n<n'$).
For $n=n'$ we found $\left[\tau_{l}(k_x)\right]_{1,1}=\left[\tau_{l}(k_x)\right]_{2,2}$ which is shown in Fig.~\ref{fig:paramtw3tww0ts1tws8}(b).

We also calculate the parameters $\left[e_l(k_x)\right]_{n,n'}$ and $\left[\tau_{l}(k_x)\right]_{n,n'}$  for 2D hosts coupled to NNN impurities. We use similar parameters as those used for NN wires without any direct coupling between the two impurities.
Similar to the NN impurities case, the diagonal terms $\left[e_l(k_x)\right]_{n,n}$ depend of the dispersion in the wire direction and the on-site chemical potentials.
For $n\neq n'$ the inter-leg hopping terms must vanish, i.e. $\left[e_l(k_x)\right]_{n,n'}=0$, since they connect sites between similar sublattices. Nevertheless, we observe finite values for these parameters after a few BL iterations  as shown in Fig.~\ref{fig:paramtw3tww0ts1tws8}(c). This is due to the loss of orthogonality in the BL calculation when we initiate from NNN impurities. However, we have observed that the accuracy is better when the separation $\vert y_a-y_b\vert$ between the two impurities is larger.
As we mentioned before, the inter-rung hopping terms $\left[\tau_{l}(k_x)\right]_{1,2}=0$ while the other values of $\left[\tau_{l}(k_x)\right]_{n,n'}$ are shown in Fig.~\ref{fig:paramtw3tww0ts1tws8}(d).

We observe that the BL iterations produce accurate results for 2D hosts coupled to NN impurities for relatively large number of iterations. Despite the less accurate results for large number of BL iterations in the case of NNN impurities we observe that at least for the minimal number of iterations (with $l=3$ corresponding to 6-leg ladders) the results are accurate enough. This will allow us to construct a minimal approximation of the full TWSS. We emphasize that this minimal approximation includes an overlap between the BL vectors generated from the two impurities and thus possibly an indirect substrate mediated coupling between wires.

\subsection{Real-space representation \label{sec:twoDladders}}
We now transform the Hamiltonian~(\ref{eq:hamiltonianBL2Imp}) back to the real-space representation in $x$-direction.
As the wire states have not been modified by the mapping of the multi-impurity subsystem to the ladder representation,
the two-wire Hamiltonian $H_{\text{w}}$ remains unchanged.
By defining new fermion operators
\begin{equation}
\label{eq:shellop}
 g^{\dag}_{xln}=\frac{1}{\sqrt{L_{x}}}\sum_{k_x}e^{-ik_xx}f^{\dag}_{k_xln}
\end{equation}
that create electrons at position $x$ in the $l$-th shell and the $n$-th 2D sheet, we get a new representation of the full wire-substrate Hamiltonian
\begin{eqnarray}
H&=&\sum^{(N_{\text {imp}}/2)}_{l=1}\sum_{x x'}\sum_{n n'}
\left[e_{l}(x-x')\right]_{nn'}g^{\dag}_{xln}g^{\phantom{\dag}}_{x'ln'} \nonumber \\
&+&\sum^{(N_{\text{imp}}/2)-1}_{l=1}\sum_{xx'}\sum_{nn'}
\left[ \left[\tau_{l}(x-x')\right]_{nn'}g^{\dag}_{xln}g^{\phantom{\dag}}_{x',l+1,n'} \right . \nonumber \\
&+& \left . \text{H.c} \right ]
\end{eqnarray}
where
\begin{equation}
 \left[e_{l}(x)\right]_{nn'}=\frac{1}{L_{x}}\sum_{k_x}\left[e_{l}(k_x)\right]_{nn'}\exp(ik_xx)
\end{equation}
are the hopping amplitudes in the wire direction within the same shell $l$
(or the on-site potential for $x=0$) whereas
\begin{equation}
 \left[\tau_{l}(x)\right]_{nn'}=\frac{1}{L_{x}}\sum_{k_x}\left[\tau_{l}(k_x)\right]_{nn'}\exp(ik_xx)
\end{equation}
are the hopping amplitudes between sites in shells $l$ and $l+1$.
Therefore, we have obtained a new representation of the Hamiltonian $H$ with long-range hoppings on two sheets of 2D lattices of size $L_x \times N_{\text{imp}}/2$.

The ladder representations of the substrate are identical for all wave vectors $k_x$ up to energy shifts.
It follows that the hopping terms between nearest-neighbor shells are
\begin{equation}
\left[\tau_{l}(x)\right]_{nn'}=-\left[t^{\text{rung}}_{l}\right]_{nn'}\delta_{x,0}
\end{equation}
with
$\left[t^{\text{rung}}_{l}\right]_{nn'}=-\left[\tau_{l}(k_x)\right]_{nn'}$.
In addition,  one finds that
\begin{equation}
\left[e_{l}(x)\right]_{nn'}=-\left[t^{\text{leg}}_x\right]_{nn'}+\left[\mu_{l}\right]_{nn'}\delta_{x,0} 
\end{equation}
with
\begin{equation}
\left[t^{\text{leg}}_x\right]_{nn'}=-\frac{1}{L_{x}}\sum_{k_x}\left[\nu(k_x)\right]_{nn'} \exp(ik_x x) 
\end{equation}
and $\left[\mu_{l}\right]_{nn'} = \left[e_{l}(k_x)\right]_{nn'} - \left[\nu(k_x)\right]_{nn'}$.

At this point, we have obtained a representation of the wire-substrate Hamiltonian $H$ in the form of two ladder-like sheets, such that each sheet has $L_x$ rungs and $N_{\text{imp}}/2$ legs, as sketched in Fig.~\ref{fig:mapping}.
The first two legs with $l=1$ are the two wires, in particular $g^{\dag}_{x,l=1,n=1} = c^{\dag}_{ax}$ and $g^{\dag}_{x,l=1,n=2} = c^{\dag}_{bx}$,
while legs with $l=2,\dots,N_{\text{imp}}/2$ correspond to the successive shells and represent the substrate.
The full Hamiltonian is made of the unchanged wire Hamiltonian $H_{\text{w}}$, the direct wire-wire coupling $H_{\text{ab}}$, the hopping terms $\left[\Gamma\right]_{nn'}$ (hybridization) between wire sites and
sites in the first two legs ($l=2$) representing the substrate, the nearest-neighbor and next-nearest-neighbor rung hoppings $\left[t^{\text{rung}}_{l}\right]_{nn'}$ between substrate legs with indices $l-1$ and $l$,
the on-site potentials and leg-leg couplings $\left[\mu_{l}\right]_{nn'}\delta_{x,0}-\left[t^{\text{leg}}_0\right]_{nn'}$ within each substrate shell, and the same intra-leg hopping terms $\left[t^{\text{leg}}_x\right]_{nn'}$ in every substrate
leg. 
The latter are identical to the hopping terms in the original substrate
Hamiltonian $H_{\text{s}}$.

For substrates with dispersions of the form
\begin{equation}
\epsilon_{\text{s}}(\mathbf{k})  = \epsilon_{\text{s}} - 2t_{\text{s}} [ \cos(k_x) + \cos(k_y) + \cos(k_z) ] 
\end{equation}
we have
$\nu(k_x)=-2t_{\text{s}} \cos(k_x)$, so that the hopping within substrate legs
takes place between nearest-neighbors only,
\begin{equation}
\left[t^{\text{leg}}_x\right]_{nn'}=\begin{cases}
                t_{\text{s}}  &\text{if } \lvert x \rvert=1 \text{ and } n=n' \,,\\
                0 &\text{otherwise}\,.
                 \end{cases}
\end{equation}
The explicit form of the full Hamiltonian is then
\begin{eqnarray}
\label{eq:ladder-hamiltonian}
 H&=&H_{\text{w}}+H_{\text{ab}}+\sum_{x,n,n'}\left[\left[\Gamma\right]_{nn'}\, g^{\dag}_{x,l=2,n}g^{\phantom{\dag}}_{x,l=1,n'}+\text{H.c.}\right]
 \nonumber \\
  &&+\sum^{N_{\text{imp}}/2}_{l=2}\sum_{x,n}\left[\mu_{l}\right]_{nn} \, g^{\dag}_{xln}g^{\phantom{\dag}}_{xln}  \nonumber \\
  &&+\sum^{N_{\text{imp}}/2}_{l=2}\sum_{x,n\neq n'}\left[\left[\mu_{l}\right]_{nn'} \, g^{\dag}_{xln}g^{\phantom{\dag}}_{xln'} + \text{H.c.}\right] \nonumber \\
  &&-t_{\text{s}} \sum^{N_{\text{imp}}/2}_{l=2}\sum_{x,n} \left [ g^{\dag}_{xln}g^{\phantom{\dag}}_{x+1,l,n} +\text{H.c.}\right] \nonumber \\
  &&-\sum^{N_{\text{imp}}/2-1}_{l=2}\sum_{x,n,n'}\left[ \left[t^{\text{rung}}_{l}\right]_{nn'} \, g^{\dag}_{xln}g^{\phantom{\dag}}_{x,l+1,n'}+\text{H.c.}\right]. 
\end{eqnarray}

For the TWSS model, 
a narrow ladder approximation (NLM) with $N_{\text{leg}}=2N_{\text{shell}}$ legs is obtained
by projecting the full Hamiltonian~(\ref{eq:ladder-hamiltonian})
onto the subspace given by the first $N_{\text{shell}}$ blocks of BL vectors, i.e. by substituting $N_{\text{shell}}\leq N_{\text {imp}}/2$ for $N_{\text {imp}}/2$ (or equivalently
$N_{\text{leg}} \leq N_{\text{imp}}$ for $N_{\text{imp}}$)
in Eq.~(\ref{eq:ladder-hamiltonian}).
As the intra-wire Hamiltonian $H_\text{w}$ is not affected by the mapping or the projection, we can
apply this procedure to systems of interacting wires ($V \neq 0$) to obtain interacting NLM.

\section{Noninteracting wires \label{sec:noninteracting}}

To compute spectral properties of the full TWSS model in the mixed representation, we use the Hamiltonian~(\ref{eq:hamiltonianWSkx}) with (\ref{eq:impurity-insulator}). 
The spectral function in this representation is given by 
\begin{equation}
\label{eq:spectrum}
A(\omega,k_x)=\sum_{\lambda=1}^{2L_yL_z+2} \delta\left(\omega-\varepsilon_{\lambda k_x}\right)
\end{equation}
where $\varepsilon_{\lambda k_x}$ $(\lambda=1,\dots,N_{\text{imp}}=2L_yLz+2)$ denote the eigenvalues of the Hamiltonians (\ref{eq:impurity-insulator}).
Similarly, to compute spectral properties of the effective NLM, we use the Hamiltonian~(\ref{eq:hamiltonianWSkx}) with $H_{k_x}$ in the BL representations~(\ref{eq:hamiltonianBL2Imp}) 
with $N_{\text{shell}} \leq N_{\text {imp}}/2$ substituted for $N_{\text {imp}}/2$.
The spectral function in this chain representation is given by 
\begin{equation}
\label{eq:spectrum2}
A(\omega,k_x)=\sum_{\lambda=1}^{N_{\text{leg}}}
\delta\left(\omega-\varepsilon_{\lambda k_x}\right)
\end{equation}
where $\varepsilon_{\lambda k_x}$ $(\lambda=1,\dots,N_{\text{leg}})$ denote the eigenvalues of these Hamiltonians. 
This spectral function can be easily calculated for any $1 \leq N_{\text{leg}} \leq N_{\text {imp}}$.

We compare spectral functions of the full system with those of the NLM with various numbers of legs. 
Unless otherwise stated, we use symmetric intra-wire hopping $t_{a}=t_{b}=t_{\text{w}}=3$ and wire-substrate hybridization $t_{a\text{c}}=t_{a\text{v}}=t_{b\text{c}}=t_{b\text{v}}=t_{\text{ws}}$.
The substrate parameters are $t_{\text{c}}=t_{\text{v}}=1$ and $\epsilon_{\text{c}}=-\epsilon_{\text{v}}=7$. The system sizes are $L_x=1000, L_y=32$ and $L_z=8$.
These model parameters correspond to an indirect gap $\Delta_{\text{s}}=2$ 
and a constant direct gap $\Delta_{\text{s}}(k_x) = 6$ for all $k_x$
in the substrate single-particle band structure [in the absence of wires or for a
vanishing wire-substrate coupling ($t_{\text{ws}}=0$)].
Our main aim in this investigation is to understand the influence of the substrate on the one-dimensional physics of the wires. Therefore, we focus on systems with two wires but without direct wire-wire coupling, i.e. we set $t_{\text{ab}}=0$. 

\begin{figure}
\includegraphics[width=0.30\textwidth]{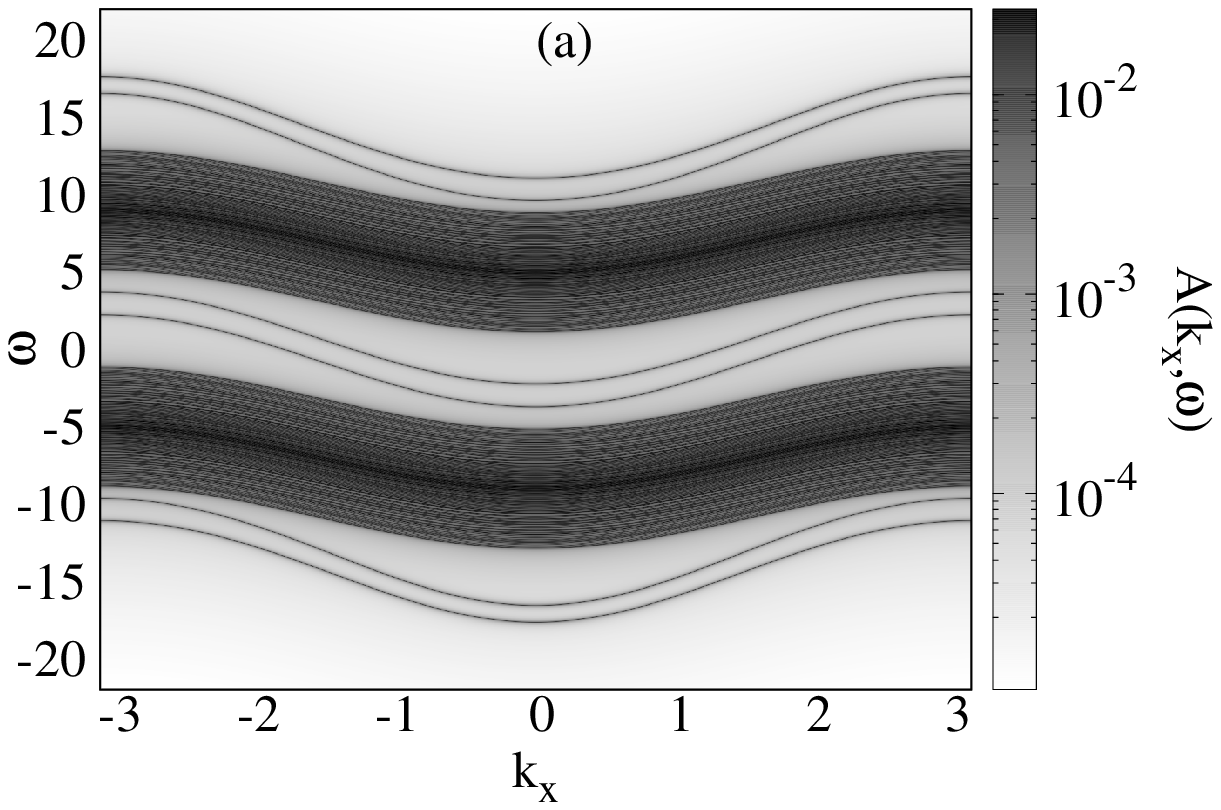}
\includegraphics[width=0.30\textwidth]{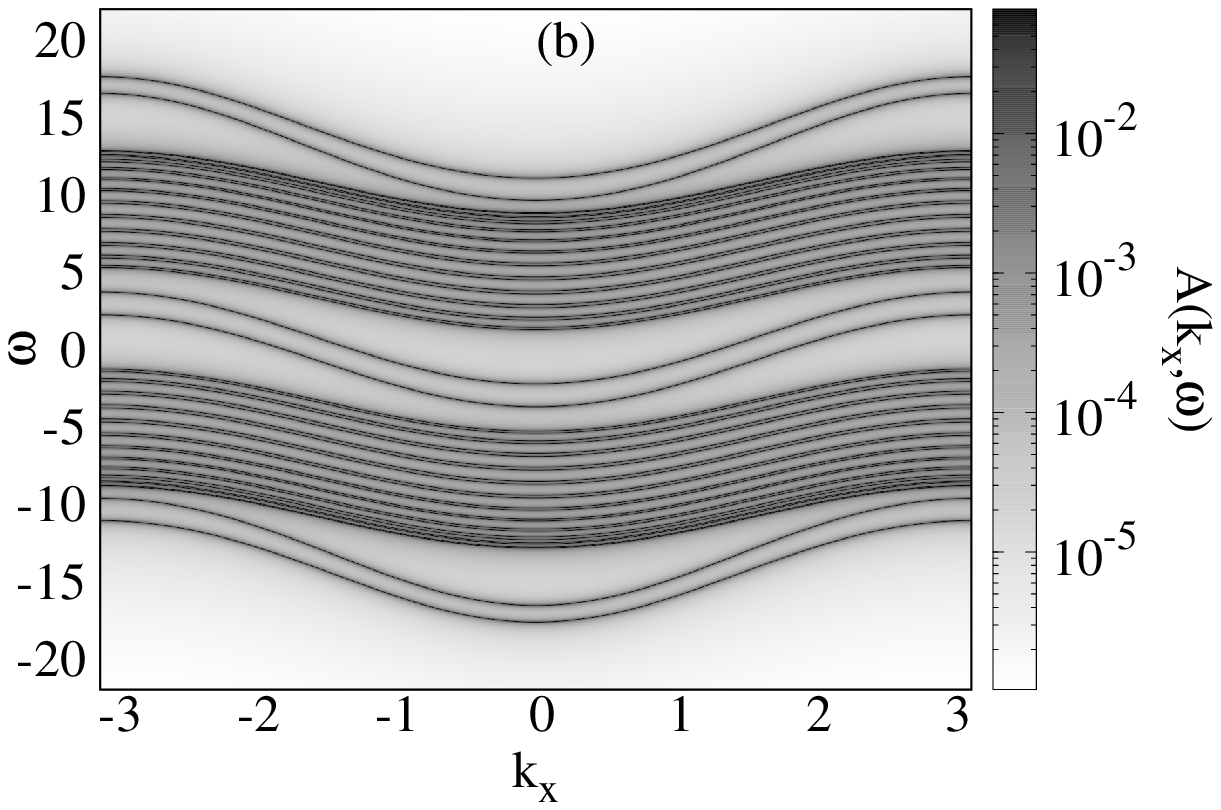}
\includegraphics[width=0.30\textwidth]{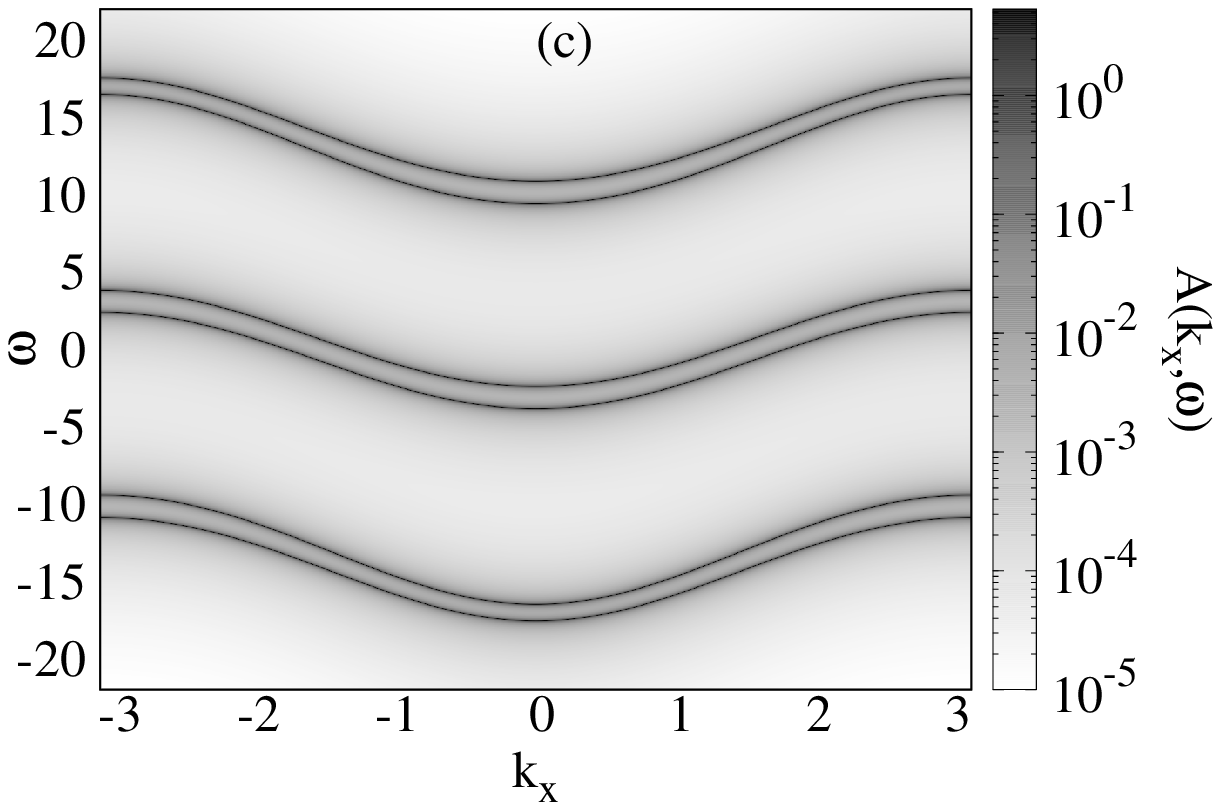}
\caption{\label{fig:NNSpect} 
Spectral functions of two NN wires on a semiconducting substrate with $t_{\text{w}}=3$, $t_{\text{ws}}=8$ and $t_{\text{s}}=1$.
(a) Full TWSS model.
(b) NLM with $N_{\text{leg}}=2N_{\text{shell}}=54$ .
(b) NLM with $N_{\text{leg}}=2N_{\text{shell}}=6$.
}
\end{figure}

Figure~\ref{fig:NNSpect}(a) displays the spectral function of the full TWSS model with NN wires and a relatively strong wire-substrate hybridization $t_{\text{ws}}=8$. Despite the absence of any direct wire-wire coupling, we clearly see two separated bands crossing the Fermi level at $\omega = \epsilon_{F} = 0$ (for half filling) in the middle of the substrate band gap. This structure corresponds to the dispersions found in a two-leg ladder
with the separation between the (bonding and anti-boding) bands given by twice the rung hopping term $t_{\perp}$ \cite{giamarchi07}.
Therefore, this observation demonstrates the existence of an effective, substrate mediated coupling between both wires even when the model
does not include the bare hopping term $t_{\text{ab}}$.

Moreover, in Fig.~\ref{fig:NNSpect} we see two bands above the substrate conduction band continuum and two other bands below the substrate valence band continuum. These four bands are due to the strong wire-substrate hybridization which forms energy levels like in a hexamer structure in first approximation $t_{\text{ws}}\gg t_{\text{w}}, t_{\text{s}}$. Each hexamer is made of two wire sites and the four substrate orbitals that are strongly hybridized to these sites by the Hamiltonian term~(\ref{eq:hybridization-insulator}). A single hexamer has six distinct energy levels and the separation increases with $t_{\text{ws}}$. Finite values of the hoppings $t_{\text{w}}$ and $t_{\text{s}}$ hybridize the energy levels of the $L_x$ hexamers in the full system and form the six bands separated from the continuum.

We see a similar behavior when we use the BL representation projected onto the subspace for $N_{\text{leg}}=2N_{\text{shell}}=54$ 
as displayed in Fig.~\ref{fig:NNSpect}(b). The two bands crossing the Fermi level resemble those seen in the full system while two other pairs of bands lie above and below 
the approximate representation of the substrate continua, respectively. However, the distribution of spectral weights in the conduction and valence bands are different from those of the full system due to the reduction of the number of bands.

We observe in Figs.~\ref{fig:NNSpect}(a) and (b) that the substrate energy gap is well approximated despite the restriction to $N_{\text{leg}} < N_{\text{imp}}$. However, by investigating different numbers of legs we find that the substrate gap increases with decreasing $N_{\text{leg}}$. This is clearly seen in Fig.~\ref{fig:NNSpect}(c) for $N_{\text{leg}}=6$. In this case, we see only six bands. Again two bands cross the Fermi level
and are similar to the two central bands observed in the full system while two other pairs of bands lie well above and below the Fermi level, respectively. These six bands were explained using the hexamer limit above but it should be noticed that in the NLM with $N_{\text{leg}}=6$
the four bands away from the Fermi energy are the remains of the two substrate continua.  
Thus these results confirm 
that a 6-leg NLM (i.e. with $N_{\text{shell}}=3$) can be a good approximation of the full system for a pair of 
NN wires 
as long as we are concerned with the physics 
occurring close to the Fermi energy $\epsilon_F=0$ on or around the wires. This agrees with and generalizes our previous findings for a single wire on a semiconducting substrate~\cite{abd17a}.

\begin{figure}
\includegraphics[width=0.30\textwidth]{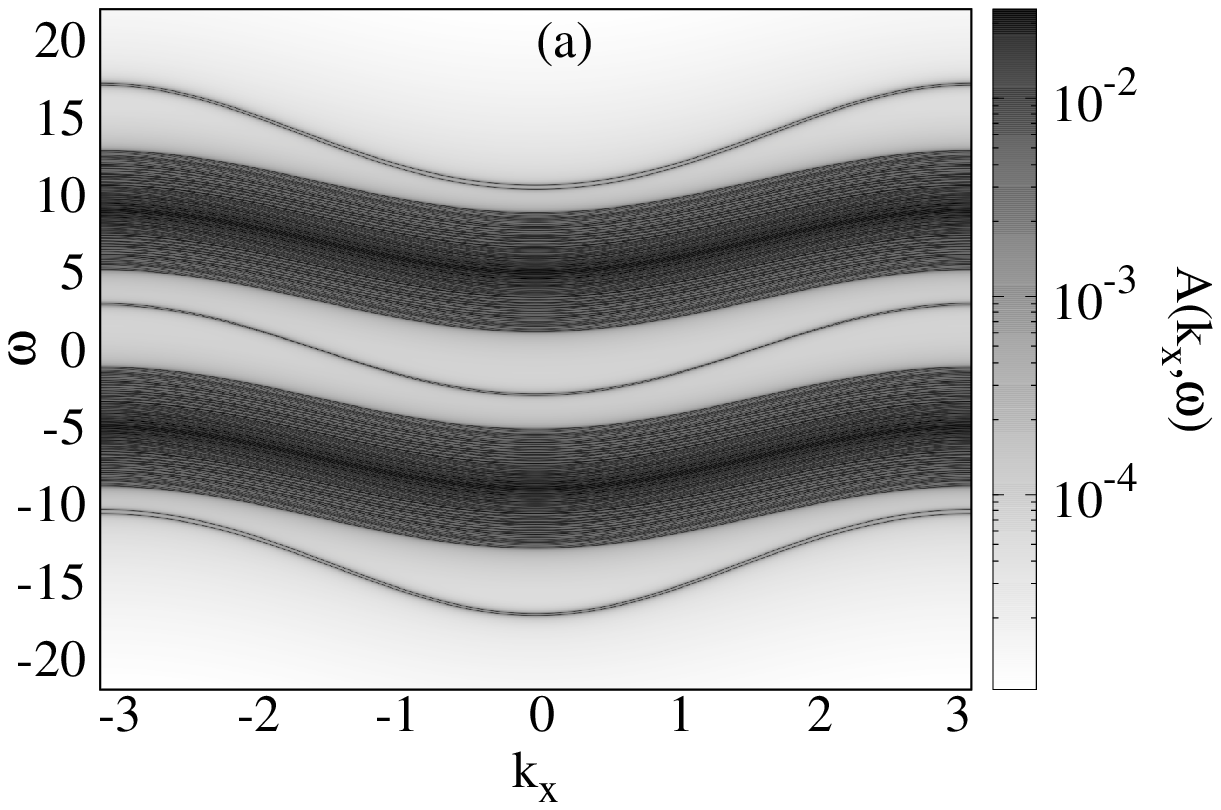}
\includegraphics[width=0.30\textwidth]{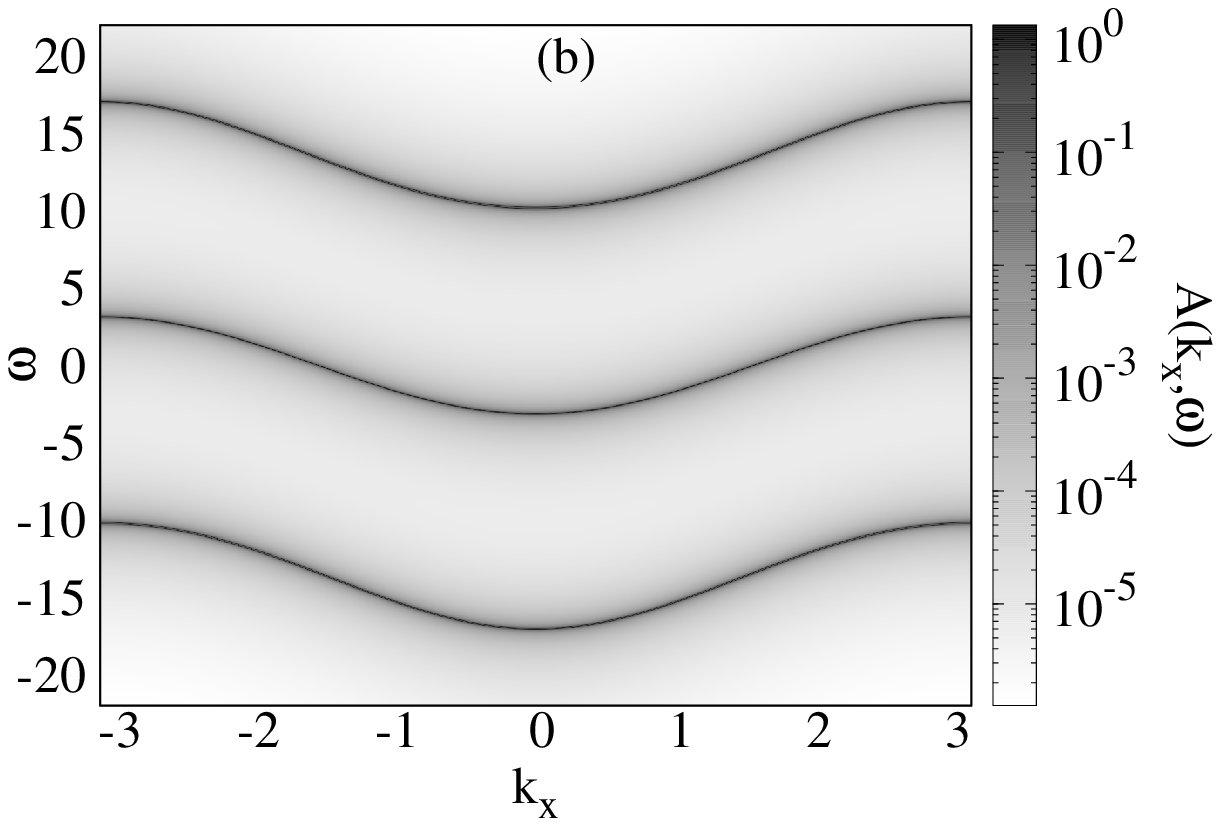}
\caption{\label{fig:NNNSpect} 
Spectral functions for two NNN wires on a semiconducting substrate with $t_{\text{w}}=3$, $t_{\text{ws}}=8$ and $t_{\text{s}}=1$.
(a) Full TWSS model.
(b) NLM with $N_{\text{leg}}=6$.
}
\end{figure}

\begin{figure}
\includegraphics[width=0.30\textwidth]{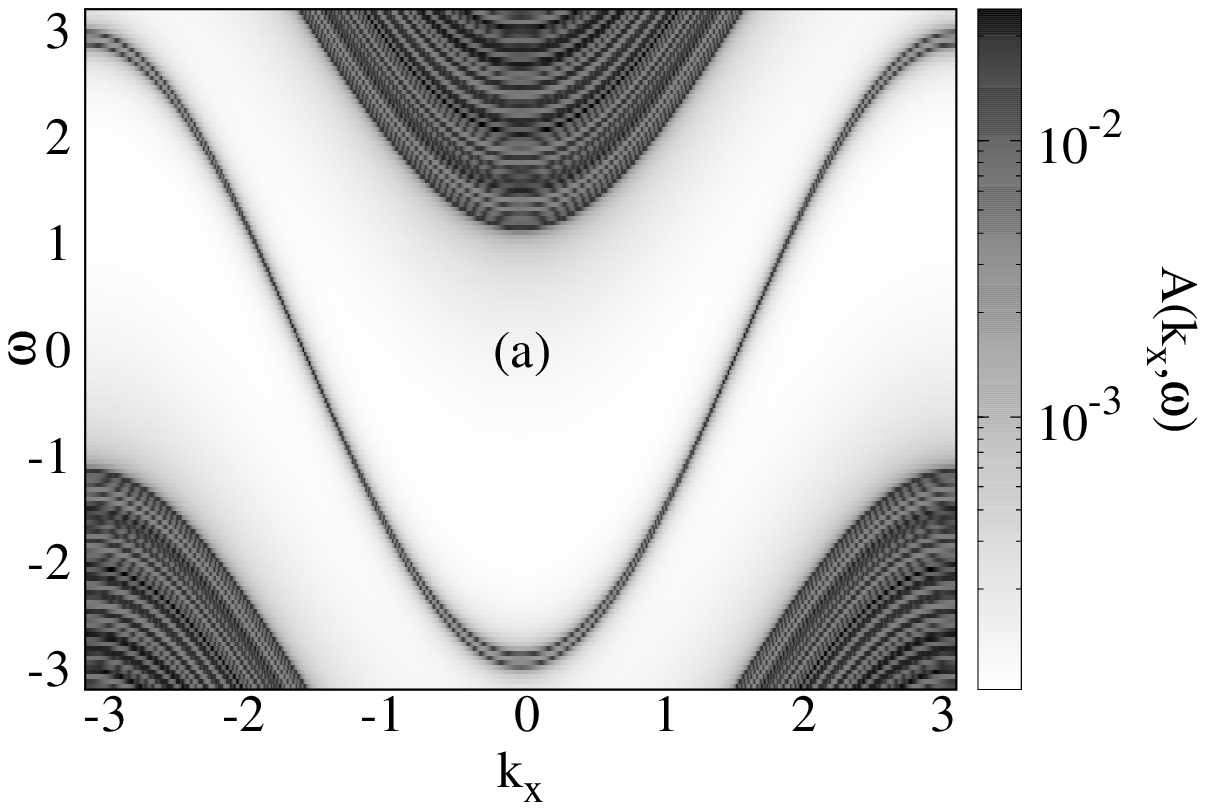}
\includegraphics[width=0.30\textwidth]{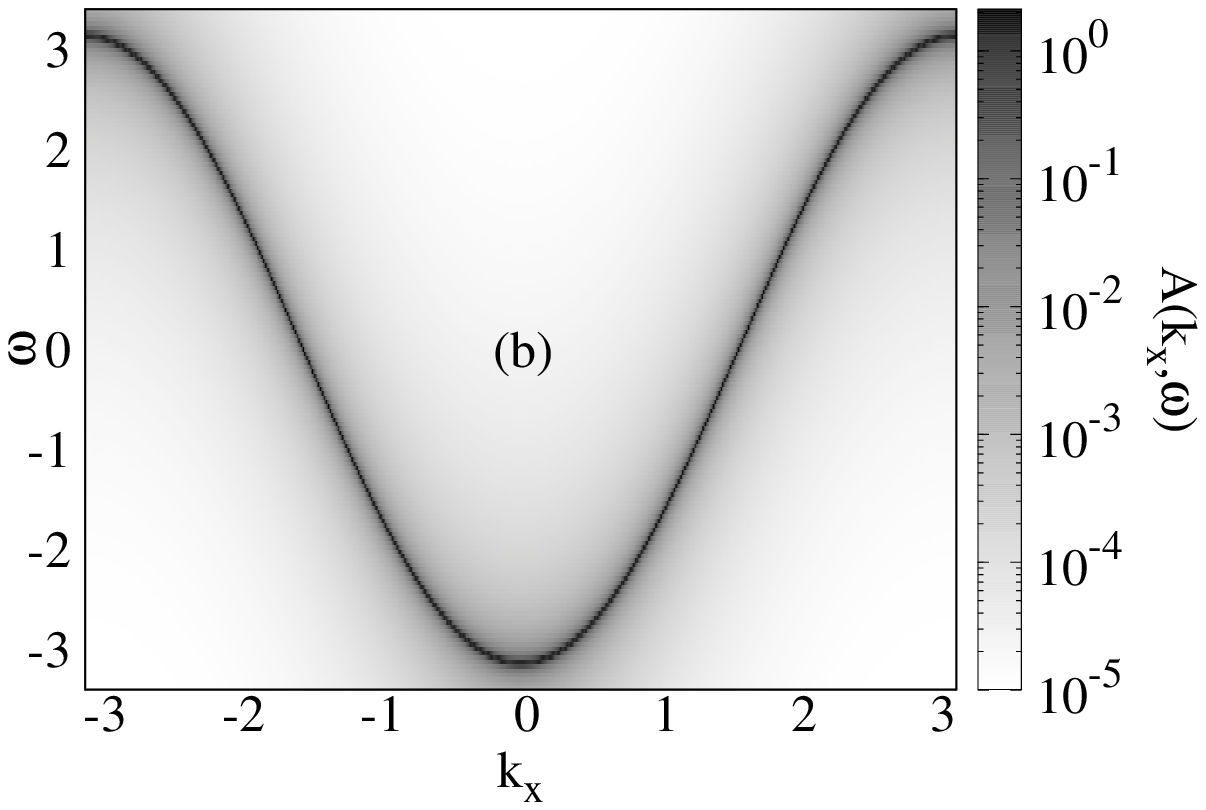}
\caption{\label{fig:NNNSpectEnlarged} 
Enlarged view of the spectral functions 
in Fig.~\ref{fig:NNNSpect}.
(a) Full TWSS model.
(b) NLM with $N_{\text{leg}}=6$.
}
\end{figure}

The full TWSS system with NNN wires reveals interesting differences in the substrate role depending on the wire positions. In Fig.~\ref{fig:NNNSpect}(a) we see three isolated dispersive features, one crossing the Fermi level inside the substrate band gap, one over the top of the conduction band continuum, and one below the bottom of the valence band continuum. Examining the central structure more closely, as shown in Fig.~\ref{fig:NNNSpectEnlarged}(a), we distinguish two bands crossing the Fermi level at $k_x= \pm \frac{\pi}{2}$ with small differences in their bandwidths.
These dispersions resemble those that would be found in a pair of uncoupled one-dimensional wires with small difference in their intra-hopping terms. Thus for NNN wires we do not find any evidence for a substrate induced hybridization 
of the two wires resulting in an effective two-leg ladder.

As we mentioned before, the BL method suffers from a fast loss of orthogonality in systems with NNN wires although it becomes more accurate for larger separations between wires. Moreover, the NLM representation generated with the  BL method for $N_{\text{shell}}=3$ does not suffer from this loss of orthogonality. We can see in Figs.~\ref{fig:NNNSpect}(b) and \ref{fig:NNNSpectEnlarged}(b) that the bands crossing the Fermi level 
are reproduced even if they are somewhat smeared out.
Therefore, we think that the 6-leg NLM ($N_{\text{shell}}=3$) could be a useful approximation of the full TWSS systems with NNN wires as in the case of NN wires. 
Contrary to the NN wires, however, the substrate does not seem to mediate an effective wire-wire coupling between NNN wires. Thus
we will investigate the effects of interaction induced correlations for these two cases.  

\section{Interacting wires \label{sec:interacting}}

In this section we investigate the 6-leg NLM approximating the TWSS model for interacting spinless fermions using the density matrix renormalization group (DMRG) method~\cite{whi92,whi93}.
Additionally, we compare with DMRG results for a two-leg ladder without substrate
 as well as for a three-leg NLM approximating a single wire on a substrate~\cite{abd18}. 
 DMRG is a well established method for quasi-one-dimensional correlated quantum lattice models~\cite{whi92,whi93,sch05,jec08a}. Recently, we have shown that DMRG can be applied to NLM with relatively large widths~\cite{abd17a,abd17b,abd18}. In this work we compute the ground-state properties of the 6-leg NLM with open boundary conditions in the leg direction as well as in the rung direction. We always simulate an even number of rungs up to $L_x=200$. The finite-size DMRG algorithm is used with up to $m=1024$ density-matrix eigenstates yielding discarded weights smaller than $10^{-5}$. We vary $m$ and extrapolate the ground-state energy to the limit of vanishing discarded weights in order to estimate the DMRG truncation error. DMRG can sometimes get stuck in metastable states in such inhomogeneous systems, hindering the convergence toward the global energy minimum. This issue cannot always be solved by increasing the number of sweeps through the lattice until convergence is reached. We have resolved this problem by targeting the lowest two eigenstates of the Hamiltonian using a single reduced density matrix for each block.
 We focus on half-filled systems, i.e. the number of spinless fermions is $N=N_{\text{leg}} \times L_x / 2$.

 Our main goal is to verify if the physics of correlated fermions in one dimension, in particular a Luttinger liquid, can occur in atomic wires on a semiconducting substrate. In a previous work~\cite{abd18}  we showed that
 the hybridization with the substrate does not preclude
 the formation of a Luttinger liquid phase in a single wire. 
 Atomic wires build regularly spaced 2D arrays of chains, however.
 It is known that a single-particle interchain hopping gives rise 
 to ordered states but this direct hopping (the parameter $t_{\text{ab}}=0$  in our model) is expected to be negligibly small in atomic wire systems
 while a Luttinger liquid can occur in the presence of interchain two-particle interactions~\cite{giamarchi07}. 
 Thus we restricted our investigation to the 6-leg NLM without direct wire-wire coupling, i.e. $t_{\text{ab}}=0$,
 and focus on two questions: (i) whether the substrate can mediate
 an effective coupling between two wires, and (ii)
 whether 1D physics, in particular a Luttinger liquid, can occur in this system.

The physics of the half-filled spinless fermion model on a two-leg ladder (i.e., our TWSS model without the substrate) was thoroughly investigated a few decades ago using field theoretical methods, renormalization group, and bosonization~\cite{giamarchi07,don01,Fabrizio1993,Yoshioka1995}
as well as exact diagonalizations~\cite{Capponi1998}. 
When the interaction is restricted to a nearest-neighbor repulsion (i.e. $V > 0$) between fermions on the same leg and the only inter-leg coupling is a single-particle rung hopping $t_{\perp} > 0$,
 the system does not have any gapless phase but is a Mott insulator
 with a charge density wave, even for arbitrarily small parameters
 $V$ and $t_{\perp}$. In contrast, an uncoupled chain ($t_{\perp}=0$)
 remains a gapless Luttinger liquid for $0 \leq V \leq 2t_{\parallel}$,
 where $t_{\parallel} > 0$ is the intra-leg hopping, while it is an insulating CDW for $V > 2t_{\parallel}$.
 The introduction of an infinitely small interchain hopping $t_{\perp}$ is sufficient to generate the Mott gap and the long-range CDW order in analytical studies~\cite{giamarchi07,don01}.
For small $t_{\perp}$ and $V$, however, gap and CDW amplitudes can be too small to be identified with certitude using DMRG due to finite-size effects. 

In a previous work~\cite{abd18} we investigated  a single spinless fermion wire on a substrate thoroughly.
We found three phases for varying couplings $V>0$:
a Luttinger liquid phase with gapless excitations
localized in the wire for $V < V_{\text{CDW}}$, a CDW insulating phase with excitations still localized in the wire for intermediate couplings $V_{\text{CDW}} < V < V_{\text{BI}}$,
and a band insulator with excitations delocalized
in the substrate for $V > V_{\text{BI}}$. An important observation is that $V_{\text{CDW}}$ increases significantly with increasing wire-substrate hybridization $t_{\text{ws}}$ above the value $V_{\text{CDW}}=2t_{\text{w}}$ for the isolated wire (i.e., in the limit $t_{\text{ws}} \rightarrow 0$). Therefore, the presence of the substrate hinders the formation of the insulating CDW ground state and stabilizes the Luttinger liquid phase.

\subsection{Single-particle gap}

\begin{figure}[t]
\includegraphics[width=0.30\textwidth]{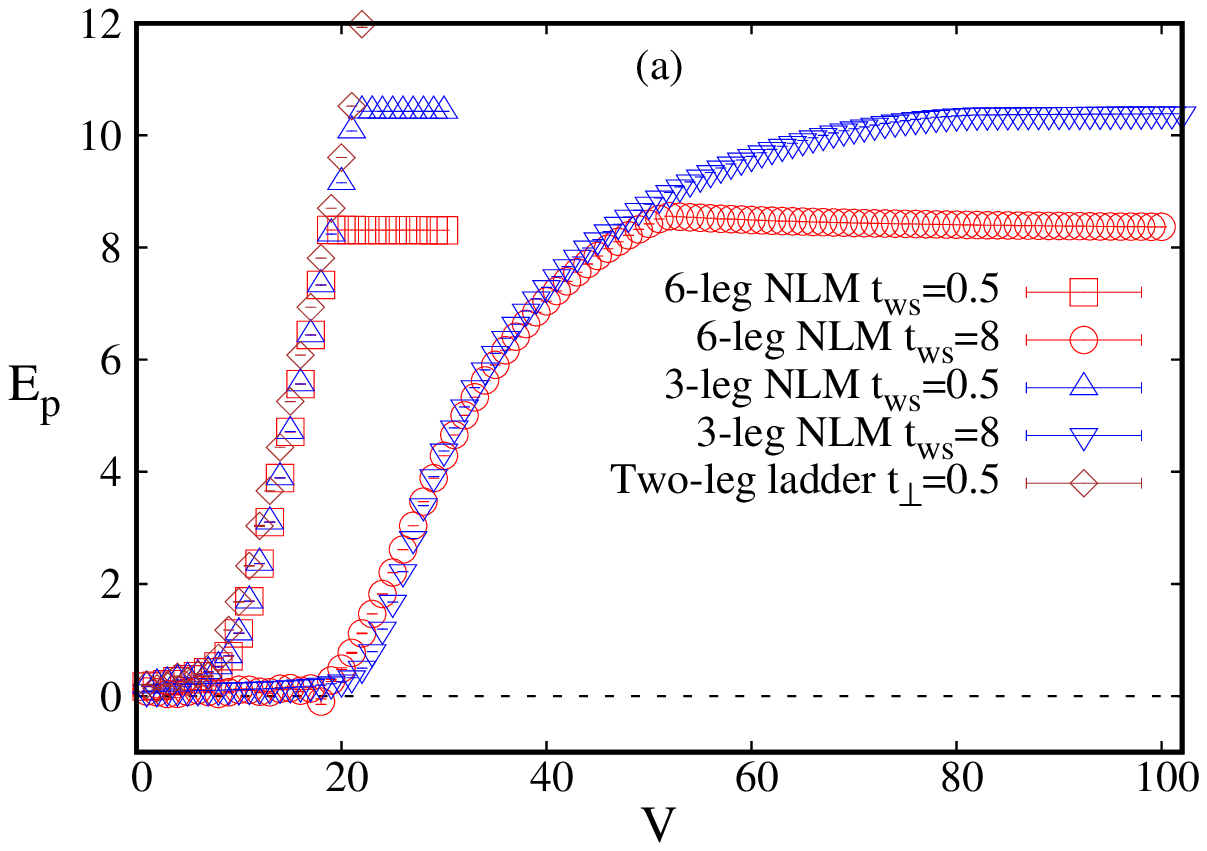}
\includegraphics[width=0.30\textwidth]{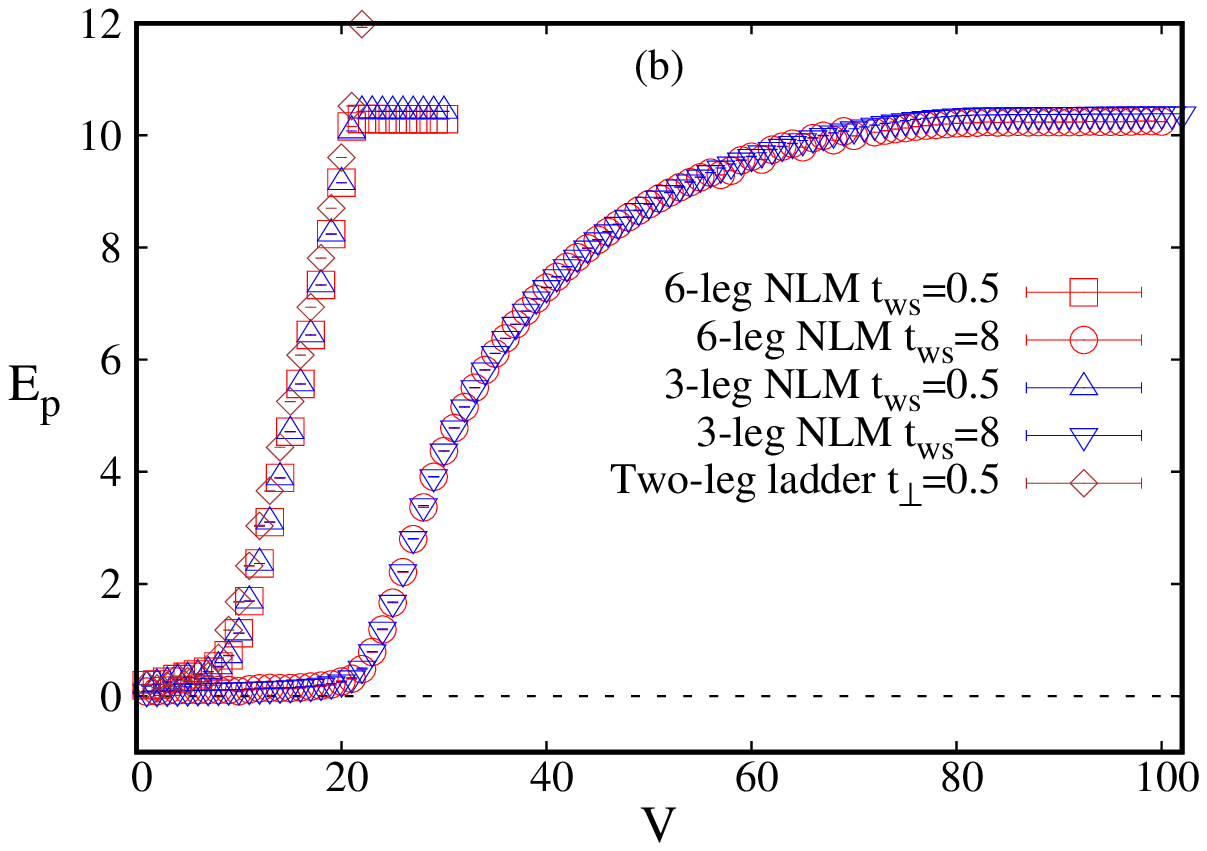}
\includegraphics[width=0.30\textwidth]{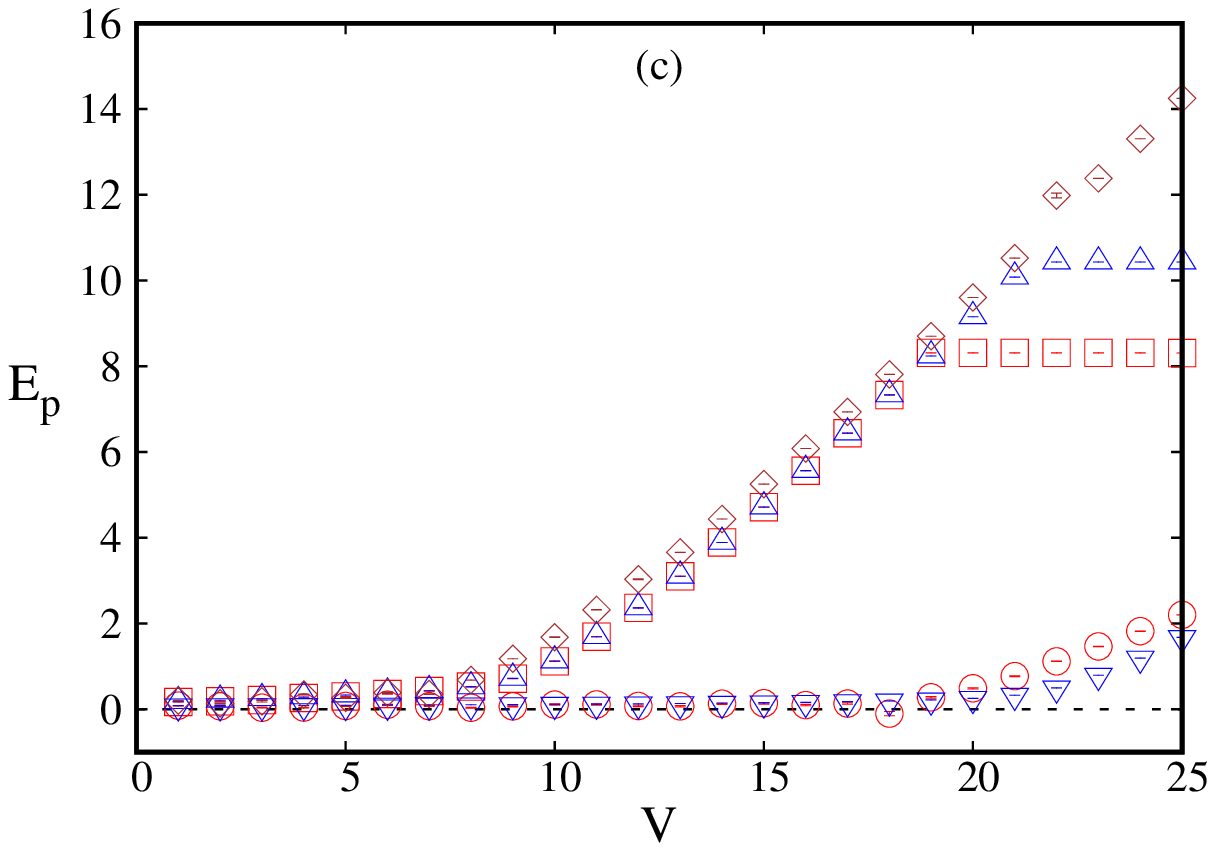}
\includegraphics[width=0.30\textwidth]{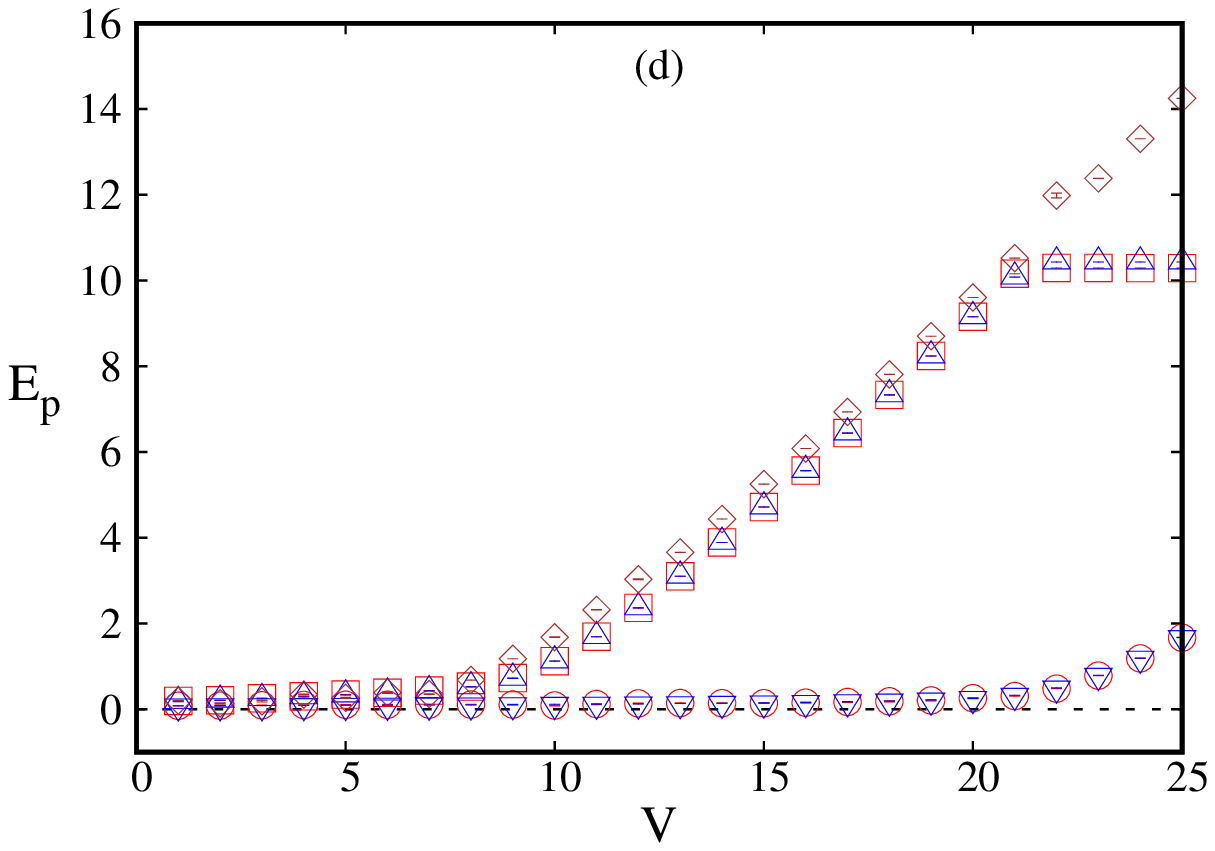}
\caption{\label{fig:Eptww0} 
Single-particle gap $E_p$ in a 6-leg NLM as function of the
intra-wire interaction $V$.
The upper plots (a) and (b) show results for 
two NN wires and two NNN wires, respectively.
The lower figures (c) and (d) show an enlarged view 
of the same results.
The upper and lower triangles are results for a three-leg NLM representing a single wire on a substrate. The diamonds 
show results for the two-leg ladder model with a rung hopping $t_{\perp}=0.5$ and a leg hopping $t_{\parallel}=3$ but no substrate.
The ladder length is $L_x=128$.
}
\end{figure}

In the light of these previous results for related systems, we now discuss the gap, the CDW order parameter,
and the density distribution of excitations in
the 6-leg NLM representation for TWSS using DMRG.
We first investigate the single-particle gap  which is defined as
\begin{equation}
E_p = E \left( N+1 \right) + E \left( N-1 \right)-2E \left( N \right)
\end{equation}
where $E(N)$ is the ground-state energy for a system with $N$ fermions.
Figure~\ref{fig:Eptww0} displays this single-particle gap as function of the interaction $V$. 
We compare the gaps for the
6-leg NLM with those for a 3-leg NLM describing a single wire on a substrate using $t_{\text{w}}=3$ and the same substrate parameters as for the  noninteracting system in the previous section. Additionally,
we show results for a two-leg ladder without substrate with a leg hopping $t_{\parallel}=3$ and a rung hopping $t_{\perp}=0.5$.

We first discussed the results for NN wires shown in Fig.~\ref{fig:Eptww0}(a). 
The two wires should be coupled
by a substrate-mediated effective hopping $t_{\perp}$
as found for noninteracting systems in the previous section.
According to the analytical findings for two-leg ladders~\cite{giamarchi07,don01}, we thus expect to observe a Mott insulator with long-range CDW order in the 6-leg NLM
for $V>0$.
For a weak wire-substrate hybridization $t_{\text{ws}}=0.5$, however, the effective coupling is weak
and the expected small single-particle gap  cannot be distinguished from finite-size effects for small $V \alt 2t_{\text{w}} = 6$ in the 6-leg NLM, see Fig.~\ref{fig:Eptww0}(c). This is similar to the finite-size gaps found in both the 3-leg NLM for a single wire
and the two-leg ladder without substrate, which are a Luttinger liquid and a Mott/CDW insulator in the thermodynamic limit for that parameter regime, respectively. 
Similarly, the single-particle gap of the 6-leg NLM  matches the Mott gap seen in the Mott/CDW phase of the 3-leg NLM for a single wire and the two-leg ladder without substrate for stronger coupling $V \gtrsim 2t_{\text{w}}$ and 
$V > 2t_{\parallel}$, respectively. This agreement persists up to the
points where the gaps of
the 6-leg and 3-leg NLM saturate (i.e. $V=V_{\text{BI}} \approx 20$ for the 3-leg NLM
with $t_{\text{ws}}=0.5$). 
This saturation marks the transition to the band insulator regime as already observed for single wires~\cite{abd18}.
The correlation gap for quasi-one-dimensional excitations in the wire increases monotonically with $V$ for all $V > V_{\text{CDW}}$. For $V > V_{\text{BI}}$, however, it becomes larger
than the NLM effective band gap for excitations delocalized in the substrate. Thus the gaps $E_p$ to the lowest single-particle excitations correspond to the effective band gaps for $V > V_{\text{BI}}$ and thus become independent from $V$.

Varying the wire-substrate hybridization up to
$t_{\text{ws}}=8$, we observe that the single-particle gaps
of the 6-leg NLM and 3-leg NLM remain very similar up to 
the saturation interaction, as shown in Fig.~\ref{fig:Eptww0}(a) for $t_{\text{ws}}=8$.
For the 3-leg NLM representing a single wire on a substrate
we know that 
a stronger hybridization $t_{\text{ws}}$ results in a smaller effective
interaction and thus in a larger critical coupling $V_{\text{CDW}}$~\cite{abd18}.
Thus, the single-particle gaps of the 3-leg NLM seen in Figs.~\ref{fig:Eptww0}(c) for $t_{\text{ws}}=8$
and $V < V_{\text{CDW}}\approx 19$ are finite-size effects while
they correspond to a finite Mott gap above this critical interaction.
The single-particle gaps of the 6-leg NLM are not significantly larger than those for the 3-leg NLM.
Therefore, we cannot determine whether the single-particle
gap of two NN wires is finite in the thermodynamic limit for all $V>0$, as expected for two-leg ladders
from the above discussion, or whether a Luttinger liquid phase
occurs at weak coupling $V$ as in a single wire on a substrate (3-leg NLM).

Figures~\ref{fig:Eptww0}(b) and (d) display $E_p$ for NNN wires in the 6-leg NLM. Again we observe a close agreement with the results for a single wire represented by the 3-leg NLM for all couplings $V$ and $t_{\text{ws}}$.
For a weak wire-substrate hybridization, these single-particle gaps are also close to the Mott gap of the two-leg ladder without substrate.
Thus, as for NN wires, we cannot determine whether
two NNN wires have a gapless Luttinger liquid phase at weak coupling or are insulating for all $V>0$.
In summary, in the regime $0<V<V_{\text{BI}}$, where a single wire on a substrate exhibits 1D physics (i.e. Luttinger liquid or Mott/CDW insulator), the analysis of the single-particle gap
does not allow us to demonstrate distinct behaviors
between NN and NNN wire pairs (e.g. like two uncoupled wires or like an effective two-leg ladder) because of finite-size effects. Thus the existence and the role of an effective
substrate-mediated wire-wire coupling remains unclear in that regime.

Remarkably, the 3D band insulator phase reveals a striking
difference between NN and NNN wires.
For NN wires the effective band gap $E_p$ in the band insulator regime (the value of $E_p$ at saturation) is significantly lower than the band gap of the 3-leg NLM for a single wire, as seen in Fig.~\ref{fig:Eptww0}(a).
For NNN wires the band gap differs only slightly from the
value found in a single-wire represented by the 3-leg NLM
as seen in Fig.~\ref{fig:Eptww0}(b).
These results demonstrate that two interacting NN wires in the 6-leg NLM
are coupled through the substrate while
the two interacting NNN wires do not feel that they share the same substrate. Therefore, the effective substrate-mediated
coupling between NN wires that we have found for noninteracting wires ($V=0$) in the previous section is confirmed at least in the band insulator regime for
strong interactions ($V > V_{\text{BI}})$.
Finally, we note that the values of $E_p$ does not change 
significantly with $t_{\text{ws}}$ in interacting NLM
although it increases with $t_{\text{ws}}$ in noninteracting NLM.

\subsection{CDW order parameter}

\begin{figure}[t]
\includegraphics[width=0.30\textwidth]{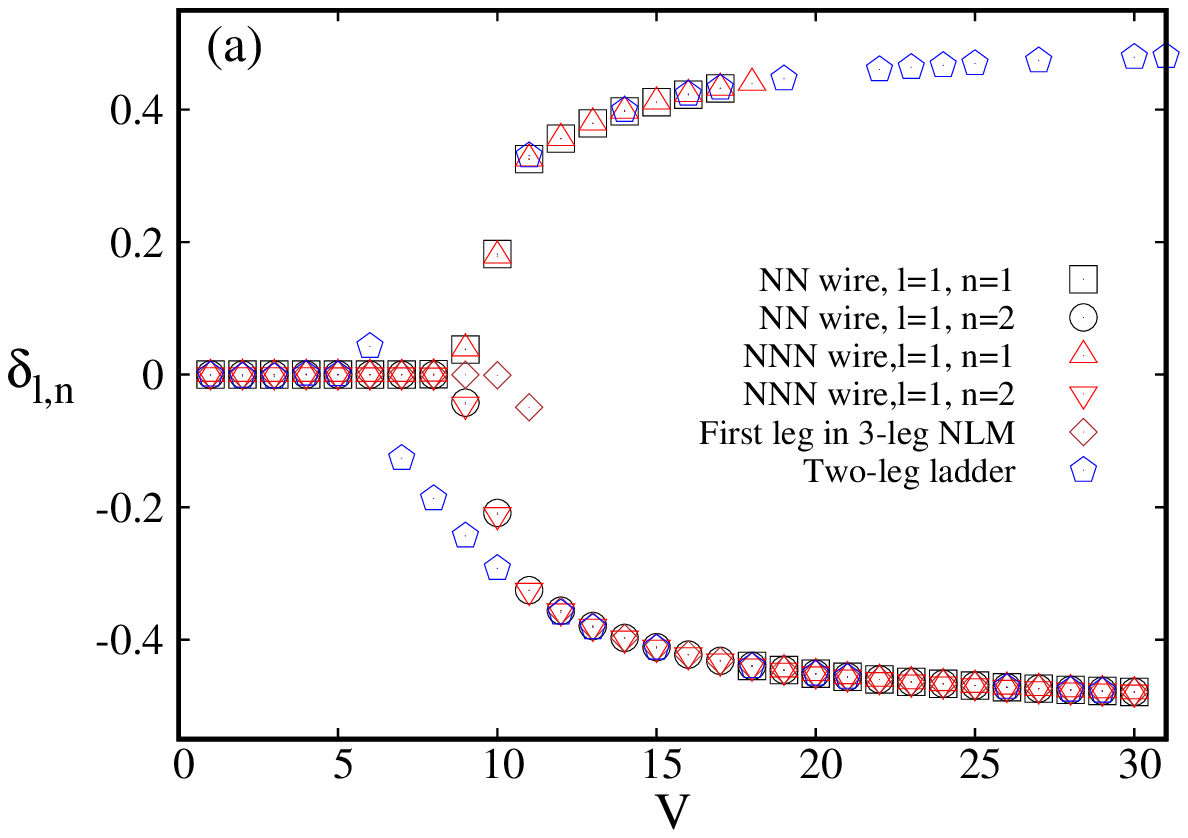}
\includegraphics[width=0.30\textwidth]{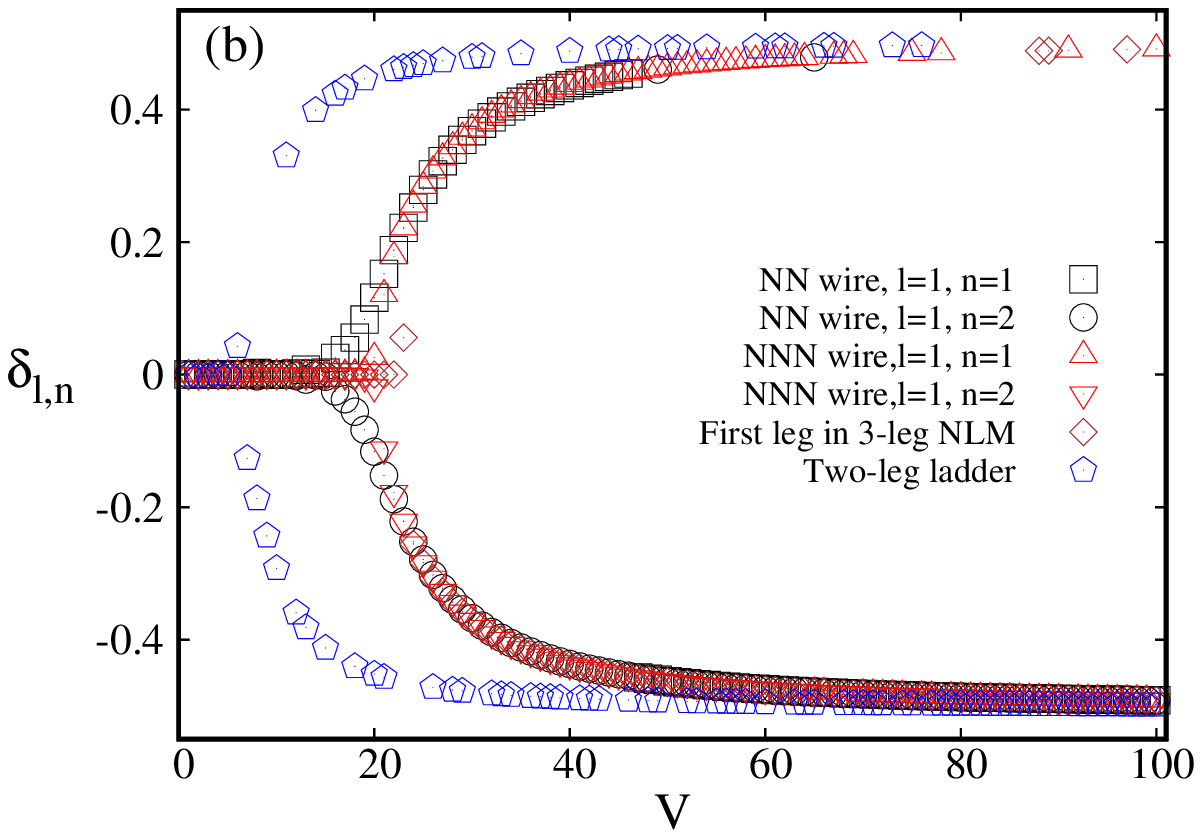}
\caption{\label{fig:CDWorderCompare} 
Charge-density-wave order parameters $\delta_{l,n}$ for various legs $(l,n)$ (see the text) in the 6-leg NLM for a pair of NN and NNN wires on a substrate.
The parameters for the original TWSS models are $t_{\text{w}}=3$, $t_{\text{ab}}=0$, $t_{\text{s}}=1$ and (a) $t_{\text{ws}}=0.5$ (b) $t_{\text{ws}}=8$.
The CDW order parameters are also shown for the first leg
of a three-leg NLM representing a single wire on a substrate as well as for the two-leg ladder model with a rung hopping $t_{\perp}=0.5$ and a leg hopping $t_{\parallel}=3$ but no substrate.
The ladder length is $L_x=128$.
}
\end{figure}

The existence of the long-range CDW order offers another way to distinguish
the Mott/CDW phase from the Luttinger liquid phase.
In the NLM the CDW order manifests itself as oscillations in the 
ground-state local density in the form
\begin{equation}
 \label{eq:CDWoscillations}
 \langle g^{\dag}_{x,l,n}g^{\phantom{\dag}}_{x,l,n} \rangle = \frac{1}{2} + (-1)^x \delta_{xln}
\end{equation}
where $\delta_{xln}$ varies slowly with $x$. 
These oscillations break the particle-hole symmetry and,
in principle, they can only occur in the thermodynamic limit. However, in DMRG calculations they become directly observable for finite systems due to symmetry-breaking truncation errors.
The CDW order parameter in each leg is thus given by
\begin{equation}
 \label{eq:CDWorderparameter}
 \delta_{l,n} = \frac{1}{L_x} \sum_{x} (-1)^x \langle g^{\dag}_{x,l,n}g^{\phantom{\dag}}_{x,l,n} \rangle .
\end{equation}

We have calculated this CDW order parameter for each leg in the 6-leg NLM using the same parameters as in the previous subsection. We again compare these results to DMRG results obtained for the 3-leg NLM representing a single wire on the substrate and for a two-leg ladder without substrate
with  $t_{\parallel}=3$ and $t_{\perp}=0.5$.
Figure~\ref{fig:CDWorderCompare} displays $\delta_{l,n}$ as a function of the 
interaction $V$. 

In Fig.~\ref{fig:CDWorderCompare}(a) we compare the results
for the two NN or NNN wires of the 6-leg NLM ($n=1,2, l=1$) and the 
single wire of the 3-leg NLM with a weak wire-substrate hybridization $t_{\text{ws}}=0.5$ as well as for the two-leg ladder without substrate. 
The order parameters  for the NN and NNN wires 
behave similarly for all interactions $V$. Moreover, they are close to the order parameter
found in the 3-leg NLM  for a single wire
(up to the arbitrary sign of $\delta_{l}$ in the single wire). A small difference is visible
around $V=10$ where the order parameters
become finite. More importantly, however, we have previously established that the transition to the CDW phase already takes
place around $V_{\text{CDW}} \approx 6$ in the 3-leg NLM with these parameters~\cite{abd18}.
Thus, although an order parameter $\delta_{l,n} \neq 0$
definitively indicates a CDW ground state,
we cannot exclude the occurrence of this CDW state
when $\delta_{l,n} \approx 0$. The CDW order parameter can be too small to be detected
with our approach. This can be seen also in the order parameters of the two-leg ladder without substrate.
The order parameters in both legs are clearly finite for $V > 2t_{\parallel} = 6$ but appears to be
vanishing below this coupling, exactly
as for a single spinless fermion chain. According to 
analytical results~\cite{giamarchi07,don01}
the ground state has a CDW long-range order
for all $V>0$, however. Due to the weak rung coupling $t_{\perp}=0.5$ the CDW order parameters are too small to be seen numerically for small interactions $V$ with our method.
We do not observe qualitative changes when 
increasing the wire-substrate hybridization 
up to $t_{\text{ws}}=8$, at least for the regime
$V<V_{\text{BI}}$, as seen in Fig.~\ref{fig:CDWorderCompare}(b).
Therefore, like for the single-particle gap,
the analysis of the CDW order
does not allow us to demonstrate distinct 1D behaviors
for NN and NNN wires or to draw a conclusion about the existence of an effective substrate-mediated coupling between wires in this parameter regime.

Again a significant difference between NN and NNN wires
become apparent in the band insulator phase 
of the 6-leg NLM.
As seen in Figs.~\ref{fig:CDWorderCompare}(a) and (b), the CDW order in the two wires
is out of phase, i.e.  $\delta_{1,1} = -\delta_{1,2}$, for $V<V_{\text{BI}}$, that is before the saturation of the single-particle gap.
In the two-leg ladder (without substrate),
it is known exactly that the ground state has
this out-of-phase configuration for all $V>0$.
In contrast, we mostly observe in-phase CDW ordering
($\delta_{1,1} = \delta_{1,2}$) for NN wires in the band insulator regime (e.g. $V>V_{BI} \approx 19$ for $t_{\text{ws}}=0.5$ and
$V>V_{BI} \approx 50$ for
$t_{\text{ws}}=8$). This stable relation between
their CDW ordering indicates strongly that each wire feels the presence of the other one in all the above cases.
However, for NNN wires
in the band insulator regime (e.g. $V>V_{BI} \approx 21$ for $t_{\text{ws}}=0.5$ and
$V>V_{BI} \approx 70$ for
$t_{\text{ws}}=8$), we observe both types 
of relative ordering indifferently.
This is seen as apparently random sign fluctuations
of $\delta_{l,n}$
for large $V$ in 
Fig.~\ref{fig:CDWorderCompare}(b).
 This observation suggests a degeneracy of the in-phase and out-of-phase CDW ordering in the NNN wires like in two uncoupled chains with CDW order (i.e., in the ground state of the two-leg ladder with $V> 2t_{\parallel}$ and $t_{\perp}=0$). 
 Therefore, the analysis of the single-particle gap
 and the CDW order parameter yields a qualitatively consistent picture for the band insulator regime
 only.
 
\subsection{Excitation density}

\begin{figure}[t]
\includegraphics[width=0.30\textwidth]{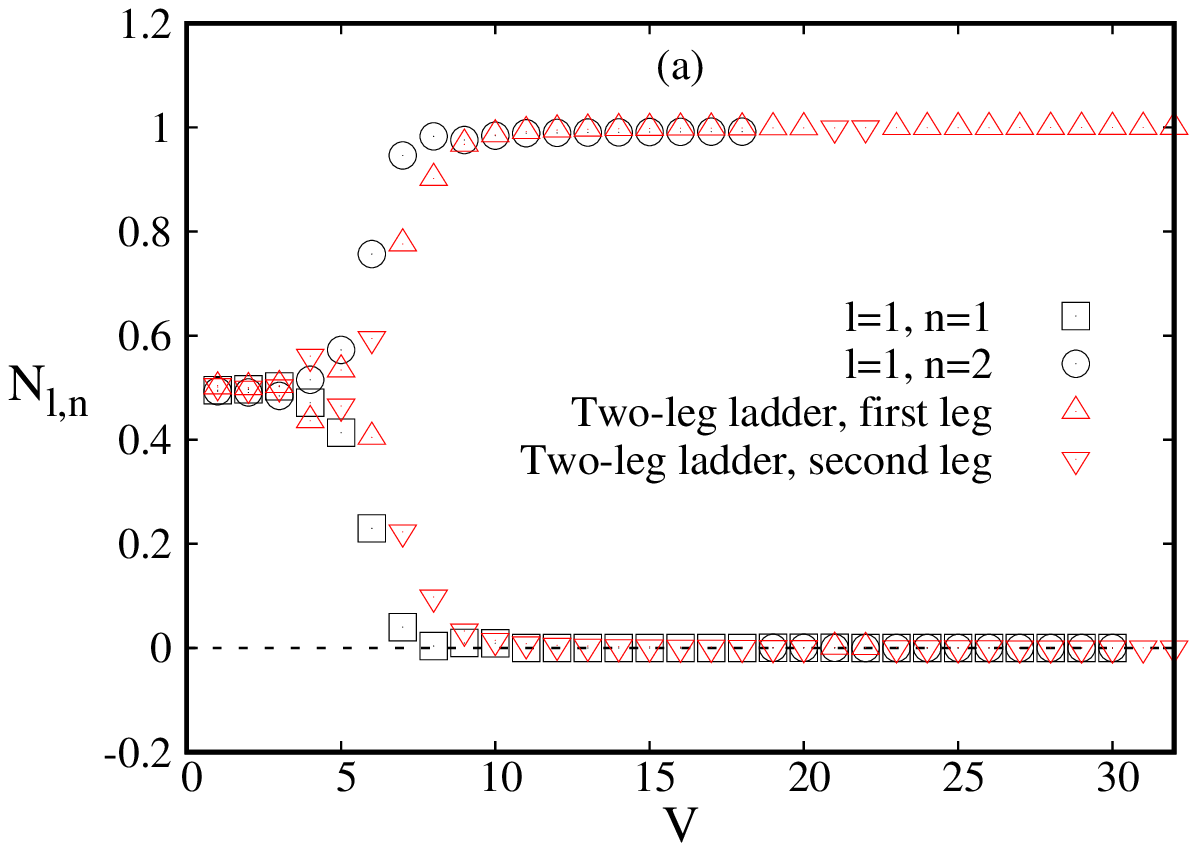}
\includegraphics[width=0.30\textwidth]{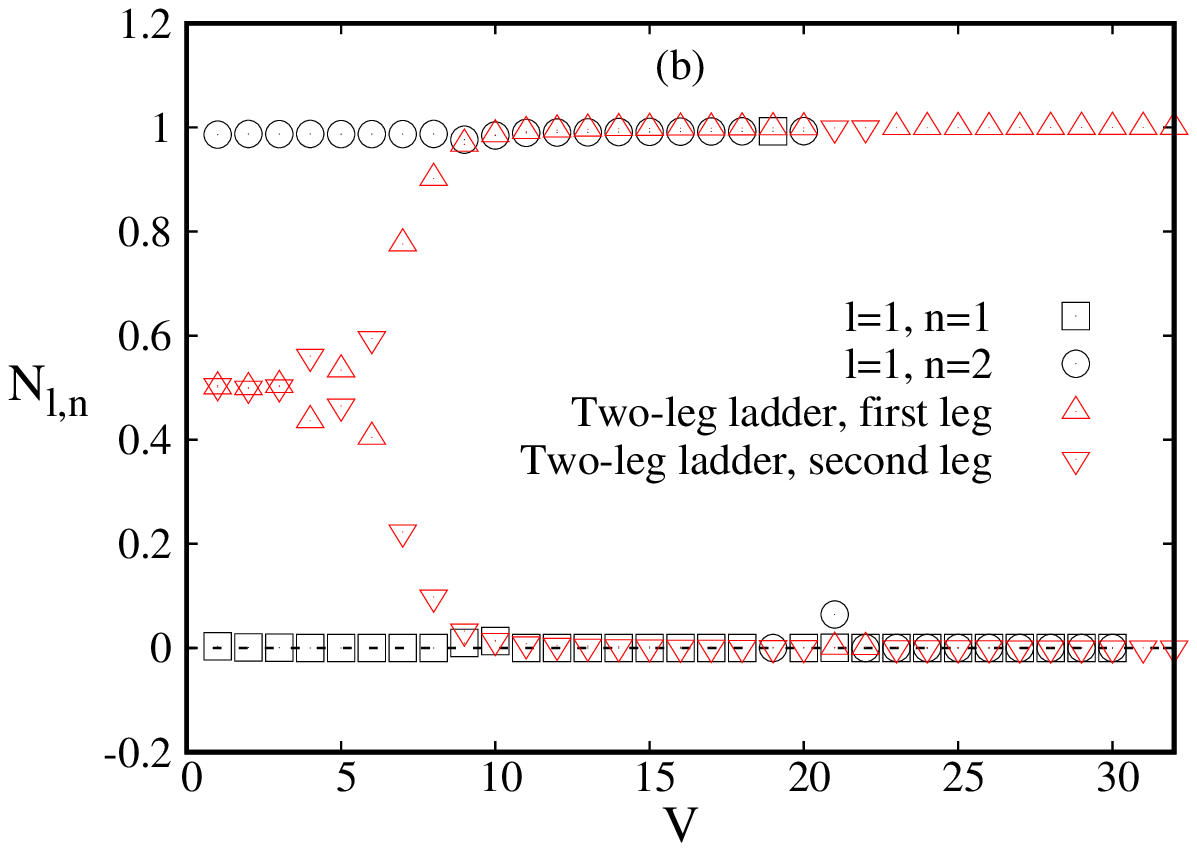}
\includegraphics[width=0.30\textwidth]{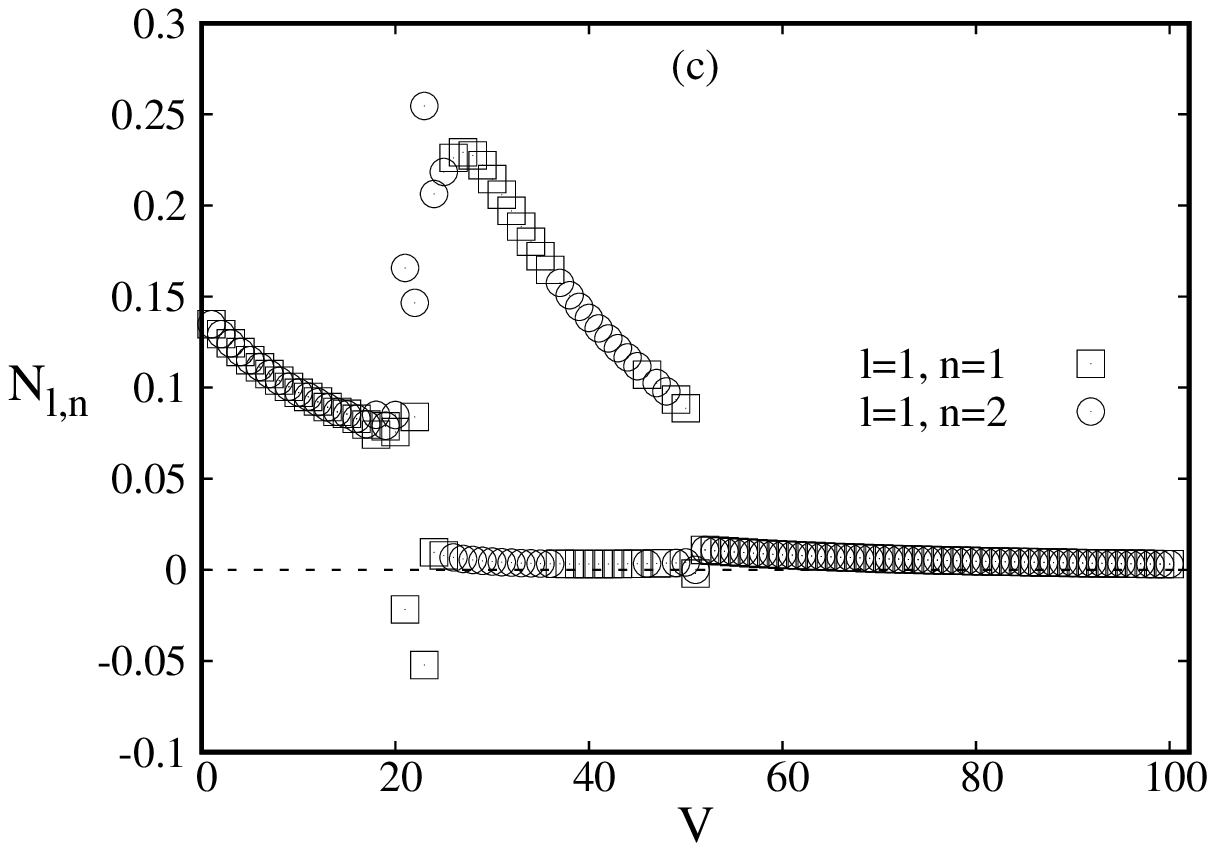}
\includegraphics[width=0.30\textwidth]{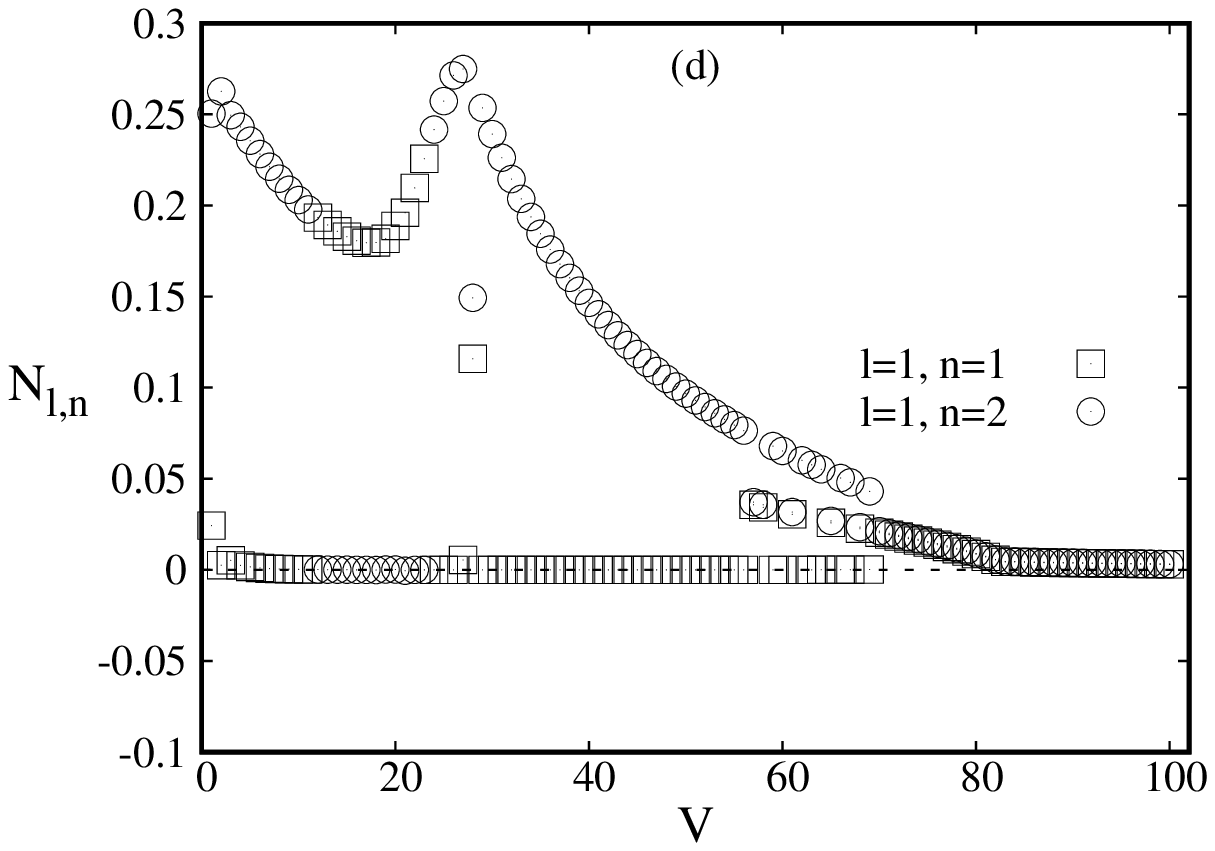}
\caption{\label{fig:chrgdens1pexCompare} 
Distribution of the single-particle excitation density $N_{l,n}$ defined in Eq.~(\ref{eq:N_ln})
between the two wires of a 6-leg NLM with $t_{\text{ws}}=3$, $t_{ab} = 0$, and $t_{\text{s}}=1$:
(a) NN wires with $t_{\text{ws}}=0.5$,
(b) NNN wires with $t_{\text{ws}}=0.5$,
(c) NN wires with $t_{\text{ws}}=8$, and
(d) NNN wires with $t_{\text{ws}}=8$.
The upper and lower triangles show results for the two-leg ladder with rung hopping $t_{\perp}=0.5$
and a leg hopping $t_{\parallel}=3$ but no substrate.
The ladder length is $L_x=128$.
}
\end{figure}

The single-particle gap and the CDW order parameters
are obvious quantities to be examined
in a system that could have Luttinger liquid and Mott/CDW insulating phases. We have seen in the previous two subsections, however, that they do not allow us to draw a conclusion
for intermediate interactions $0< V < V_{\text{BI}}$.
Thus we now turn to the density distribution of low-energy
excitations to gain more information.
Similar quantities have already proven to be useful
to understand the ground state of inhomogeneous ladder systems~\cite{abd17b,abd18,abd15,Essalah21}.

The distribution of single-particle excitations in the legs provides the 
clearest evidence for the difference between NN and NNN wires in the 6-leg NLM.
This distribution is defined as the variation of the total density in each leg
of the NLM when a fermion is added to the half-filled system
\begin{equation}
\label{eq:N_ln}
N_{l,n} = \sum_x \langle g^{\dag}_{x,l,n}g^{\phantom{\dag}}_{x,l,n} \rangle - \frac{L_x}{2},
\end{equation}
where the expectation value is calculated for the ground state with $N = (N_{\text{leg}} \times L_x/2)+1$ particles.
The evolution of $N_{l,n}$ in the wires (i.e. for $l=1$) is displayed in Fig.~\ref{fig:chrgdens1pexCompare}
for increasing interaction strength $V$.
Note that $N_{1,n}$ vanishes
in the band insulating regime ($V > V_{\text{BI}})$
because low-energy excitations are delocalized
in (the wires representing) the substrate. Actually, this is how we can determine $V_{\text{BI}}$ accurately.

For $V < V_{\text{BI}}$ and a weak hybridization $t_{\text{ws}}=0.5$ 
we observe 
in Fig.~\ref{fig:chrgdens1pexCompare}(a) that
the excitation density distribution in the NN wires
of the 6-leg NLM is very similar to the distribution found
in the two-leg ladder without substrate.
Excitations are distributed equally in both wires for weak interactions $V$ but become localized in one wire
for stronger interactions.
A different behavior is found in the 6-leg NLM for NNN wires with the same parameters, as 
shown in Fig.~\ref{fig:chrgdens1pexCompare}(b).
In that case the  single-particle excitations are entirely localized 
in one wire starting from the smallest value of $V$.
The difference between NN and NNN wires remains qualitatively similar in systems with
stronger wire-substrate hybridization $t_{\text{ws}}$.
This is illustrated in Figs.~\ref{fig:chrgdens1pexCompare} (c) and (d) for $t_{\text{ws}}=8$.
The main change for stronger $t_{\text{ws}}$ is that an increasing fraction
of the density is distributed in the substrate shells. Thus the difference between both
types of excitations becomes less striking for $N_{1,n}$ alone.

A localization of excitations in the wires (for weak 
$t_{\text{ws}}$ or in the shells around each wire
for strong $t_{\text{ws}}$) is similar to our findings
for a single wire on a substrate~\cite{abd18}.
It confirms the 1D nature of the wires in the TWSS model
represented by the 6-leg NLM for $V < V_{\text{BI}}$.
In addition, the behavior of low-energy excitations is similar in the NN wires and in the two-leg ladder with a finite rung hopping $t_{\perp}$
and thus shows that the NN wires are effectively coupled. In contrast, low-energy excitations of NNN wires
behave like in two uncoupled chains,
e.g. in the two-leg ladder with $t_{\perp} \rightarrow 0$.

Therefore, the analysis of the excitation density 
shows that NN wires in the 6-leg NLM are coupled by an effective substrate-mediated hybridization
while NNN wires remain uncoupled when the interaction
$V$ is finite but not too strong.
This complements our findings for noninteracting
systems ($V=0$) in the previous section
and for the band insulator regime ($V > V_{\text{BI}}$)
in the previous two subsections.

\section{\label{sec:conclusion} Conclusion}

In previous works we showed how to map models for
a single correlated quantum wire deposited on an insulating substrate onto narrow ladder models (NLM) that can be studied 
with the DMRG method~\cite{abd17a}. We used this approach to show
that the 1D Luttinger liquid and CDW insulating phases
found in isolated spinless fermion chains
can survive the coupling to a substrate~\cite{abd18}.
In this work we have extended this mapping to 
multi wires on a substrate using a block Lanczos algorithm. A minimal 6-leg NLM has been used  successfully to approximate a system
made of two wires on a semiconducting substrate (TWSS)  but numerical errors originating from the loss of orthogonality of block Lanczos vectors could be
an issue for broader ladders.
 
We have applied this approach to an interacting spinless fermion model for two wires on a substrate. Studying the resulting 6-leg NLM without
a direct coupling between wires,
we have found that low-energy single-particle excitations are localized
in or around the wires for weak to intermediate interactions $V$. Thus the TWSS realizes an
effectively 1D correlated system
in agreement with the previous detailed study 
of a single wire on a substrate~\cite{abd18}.

The main result of the present study is the discovery of the nonuniversal influence of the 
substrate on the effective ladder system built by the two wires in the 6-leg NLM. We have found that the two wires are coupled by an effective substrate-mediated hybridization
when both wires are deposited on top of nearest-neighbor (NN) sites of the substrate lattice but not when they are
positioned on top of next-nearest-neighbor (NNN) sites.
More generally, for noninteracting wires we observe 
a substrate-mediated coupling when adjacent wire sites belong to different sublattices of the bipartite  lattice but not when they belong to the same  sublattice.

In the absence of a direct wire-wire coupling (as expected in real atomic wire systems), the substrate-mediated 
effective coupling could then have a decisive influence on the
wire properties.
According to analytical results for a two-leg spinless fermion ladder (without substrate), 
the 6-leg NLM with NN wires should be a Mott insulator with long-range
CDW order for any coupling $V>0$ while for NNN wires it should
be Luttinger liquid for weak interactions up to a finite critical value $V_{\text{CDW}}$ like
for uncoupled wires.
Unfortunately, we could not verify directly that the single-particle gap
or the CDW ordering are different for the NN and NNN wires. Both gaps and CDW amplitudes are very small
for weak interactions $V$ and weak wire-wire coupling.
Due to the high cost of DMRG computations
for the 6-leg NLM, we have not been able distinguish
them from finite-size effects. 
Nevertheless, the density distributions of single-particle excitations between the two wires indicate clearly an effective coupling between NN wires but effectively decoupled NNN wires.
Therefore, we cannot determine whether the Luttinger liquid phase of isolated wires can
exist when more than one wire is deposited on a substrate or the substrate-mediated coupling
leads 
systematically  to insulating ground states like the Mott/CDW phase. 

We think that this question could be resolved in the future for the 6-leg NLM model constructed in this work
using other methods for 1D correlated systems.
For instance bosonization and renormalization group methods can probably access parameter regimes that are out of reach using DMRG.
A question that we could not address in the present work is whether broader ladder approximations 
of the wire-substrate systems could lead to different conclusions.
The previous systematic studies of single wires 
on substrates suggest that the results found in the minimal NLM (6 legs) should remain at least qualitatively valid
for broader NLM.
Other numerical approaches such as quantum Monte Carlo can treat not only broader NLM  but also
the three-dimensional wire-substrate model directly
and thus can be used to complement the NLM approach~\cite{abd17a,abd17b}.

Finally, our investigations show that it may be difficult to determine under which conditions the physics of correlated one-dimensional electrons can be realized in arrays of atomic wires on semiconducting substrates because they seem to 
depend on the model (and consequently material) particulars.

\acknowledgments

We would like to thank T. Shirakawa for fruitful discussions on the BL algorithm.
This work was done as part of the Research Units Metallic nanowires on the atomic scale: Electronic and vibrational coupling in real world systems (FOR1700) of the German Research Foundation (DFG) and was supported by grant JE 261/1-2. The DMRG calculations were carried out on the cluster system at the Leibniz University of Hannover.


\end{document}